\documentclass[twocolumn,tighten,twocolappendix]{aastex631}
\usepackage{wrapfig}
\usepackage{graphics,graphicx}
\usepackage{epstopdf}
\usepackage[latin1,utf8]{inputenc}
\usepackage{tikz}
\usepackage{booktabs}
\usetikzlibrary{shapes,arrows}
\usepackage{amssymb, amsmath, amsthm}
\usepackage[normalem]{ulem}
\usepackage{soul} 
\usepackage{float}
\usepackage{color}
\usepackage{xcolor}
\definecolor{ultramarine}{rgb}{0.01, 0.64, 0.86} 
\def\green#1 {{\textcolor{ultramarine}{#1}}\ }

\newcommand{\msun}{$ M_\odot$}

\newcommand{\jybeam}{Jy\,beam$^{-1}$}
\newcommand{\mjybeam}{mJy\,beam$^{-1}$}

\newcommand{\parcsec}{\mbox{$.\!\!\arcsec$}}
\newcommand{\ssstyle}{\scriptscriptstyle}

\shorttitle{Decoding SFR versus $M_{\rm dense}$ relation}

\begin{document}

\title{Why is the Star Formation Rate Proportional to Dense Gas Mass?}


\author{Sihan Jiao}
\affil{
National Astronomical Observatories, Chinese Academy of Sciences, 20A Datun Road, Chaoyang District, Beijing 100012, China
}

\author[0000-0001-5950-1932]{Fengwei Xu}
\affil{Kavli Institute for Astronomy and Astrophysics, Peking University, Beijing 100871, People's Republic of China}
\affil{Department of Astronomy, School of Physics, Peking University, Beijing, 100871, People's Republic of China}
\affil{I. Physikalisches Institut, Universität zu Köln, Zülpicher Str. 77, D-50937 Köln, Germany}

\author[0000-0003-2300-2626]{Hauyu Baobab Liu}
\affil{Department of Physics, National Sun Yat-Sen University, No. 70, Lien-Hai Road, Kaohsiung City 80424, Taiwan, R.O.C.}
\affil{Center of Astronomy and Gravitation, National Taiwan Normal University, Taipei 116, Taiwan}

\author{Yuxin Lin}
\affil{Max-Planck-Institut f\"ur Extraterrestrische Physik, Giessenbachstr. 1, D-85748 Garching bei M\"unchen, Germany}

\author{Jingwen Wu}
\affil{University of Chinese Academy of Sciences, Beijing 100049, China}
\affil{
National Astronomical Observatories, Chinese Academy of Sciences, 20A Datun Road, Chaoyang District, Beijing 100012, China
}

\author{Zhi-Yu Zhang}
\affiliation{School of Astronomy and Space Science, Nanjing University, Nanjing 210093, China}
\affiliation{Key Laboratory of Modern Astronomy and Astrophysics, Ministry of Education, Nanjing 210093, China}

\author{Zhiqiang Yan}
\affiliation{School of Astronomy and Space Science, Nanjing University, Nanjing 210093, China}
\affiliation{Key Laboratory of Modern Astronomy and Astrophysics, Ministry of Education, Nanjing 210093, China}

\author{Di Li}
\affil{New Cornerstone Science Laboratory, Department of Astronomy, Tsinghua University, Beijing 100084, China}
\affil{
National Astronomical Observatories, Chinese Academy of Sciences, 20A Datun Road, Chaoyang District, Beijing 100012, China
}

\author[0000-0002-9390-9672]{Chao-Wei Tsai} 
\affil{
National Astronomical Observatories, Chinese Academy of Sciences, 20A Datun Road, Chaoyang District, Beijing 100012, China
}
\affil{Institute for Frontiers in Astronomy and Astrophysics, Beijing Normal University,  Beijing 102206, China}
\affil{University of Chinese Academy of Sciences, Beijing 100049, China}

\author{Yongkun Zhang}
\affil{
National Astronomical Observatories, Chinese Academy of Sciences, 20A Datun Road, Chaoyang District, Beijing 100012, China
}
\affil{University of Chinese Academy of Sciences, Beijing 100049, China}

\author{Linjing Feng}
\affil{
National Astronomical Observatories, Chinese Academy of Sciences, 20A Datun Road, Chaoyang District, Beijing 100012, China
}
\affil{University of Chinese Academy of Sciences, Beijing 100049, China}

\author[0000-0002-7237-3856]{Ke Wang}
\affil{Kavli Institute for Astronomy and Astrophysics, Peking University, Beijing 100871, People's Republic of China}

\author{Zheng Zheng}
\affil{
National Astronomical Observatories, Chinese Academy of Sciences, 20A Datun Road, Chaoyang District, Beijing 100012, China
}
\affil{University of Chinese Academy of Sciences, Beijing 100049, China}

\author{Fanyi Meng}
\affil{Department of Astronomy, Tsinghua University, Beijing 100084, People’s Republic of China}
\affil{University of Chinese Academy of Sciences, Beijing 100049, China}

\author{Hao Ruan}
\affil{University of Chinese Academy of Sciences, Beijing 100049, China}
\affil{
National Astronomical Observatories, Chinese Academy of Sciences, 20A Datun Road, Chaoyang District, Beijing 100012, China
}

\author{Fangyuan Deng}
\affil{University of Chinese Academy of Sciences, Beijing 100049, China}
\affil{
National Astronomical Observatories, Chinese Academy of Sciences, 20A Datun Road, Chaoyang District, Beijing 100012, China
}

\author[0009-0003-2243-7983]{Keyun Su}
\affiliation{Kavli Institute for Astronomy and Astrophysics, Peking University, Beijing 100871, People's Republic of China}

\begin{abstract}
One of the most profound empirical laws of star formation is the Gao-Solomon relation, a linear correlation between the star formation rate (SFR) and the dense molecular gas mass. It is puzzling how the complicated physics in star-formation results in this surprisingly simple proportionality. Using archival {\it Herschel} and Atacama Large Millimeter/submillimeter Array Observations, we derived the masses of the most massive cores ($M^{\rm max}_{\rm core}$) and masses of the gravitationally bound gas ($ M_{\rm gas}^{\rm bound}$) in the parent molecular clouds for a sample of low-mass and high-mass star-forming regions. We discovered a significant correlation $\log(M^{\rm max}_{\rm core}/M_{\odot}) = 0.506 \log(M_{\rm gas}^{\rm bound}/M_{\odot})-0.32$. 
Our discovered $M^{\rm max}_{\rm core}$-$M_{\rm gas}^{\rm bound}$ correlation can be approximately converted to the Gao-Solomon relation if there is (1) a constant 30\% efficiency of converting $M^{\rm max}_{\rm core}$ to the mass of the most massive star ($m^{\rm max}_{\rm star}$), and (2) if SFR and $m^{\rm max}_{\rm star}$ are tightly related through $\log({\rm SFR}/(M_{\odot} {\rm yr}^{-1})) = 2.04 \log(m^{\rm max}_{\rm star}/M_{\odot})-5.80$.
Intriguingly, both requirements have been suggested by previous theoretical studies (c.f. \citealt{Yan2017A&A...607A.126Y}).
Based on this result, we hypothesize that the Gao-Solomon relation is a consequence of combining the following three non-trivial relations (i) SFR vs. $m^{\rm max}_{\rm star}$, (ii) $m^{\rm max}_{\rm star}$ vs. $M^{\rm max}_{\rm core}$, and (iii) $M^{\rm max}_{\rm core}$ vs. $M_{\rm gas}^{\rm bound}$. This finding may open a new possibility to understand the Gao-Solomon relation in an analytic sense.
\end{abstract}
\keywords{Star formation(1569) --- Initial mass function(796) --- Molecular clouds(1072)}

\section{Introduction}\label{intro}

The Gao-Solomon relation \citep[][]{Gao2004}
\begin{equation}
{\rm SFR} = 1.8 \times 10^{-8} \left(\frac{M^{\rm dense}_{\rm gas}}{1 \, M_{\odot}}\right) \, M_{\odot} \, \rm yr^{-1},
\end{equation}
was first discovered in the spatially unresolved surveys of the star-formation rates (SFR) and the masses of the dense molecular gas reservoir ($M^{\rm dense}_{\rm gas}$)\footnote{We note that the ordinary Gao-Solomon relation was a correlation between the HCN luminosity and the infrared luminosity. 
In the earlier works, HCN was regarded a tracer of high-density molecular gas (e.g., \citealt{Evans1999ARA&A..37..311E}).
However, recent surveys of star-forming regions suggested that a significant fraction of the HCN J$=$1–0 emission originates from low-density regions \citep[e.g.,][]{Pety2017,Kauffmann2017,Evans2020ApJ,Dame2023ApJ}.
In these low-density regions, HCN emissions can be excited by electron collisions, allowing them to trace UV photons and feedback processes rather than dense gas \citep[][]{Goldsmith2017ApJ,Kauffmann2017,Goicoechea2022,Jones2023MNRAS,Santa-Maria2023}.
Therefore, the ordinary Gao-Solomon relation may be alternatively explained by feedback-driven HCN emission instead of the correlation between $M_{\rm gas}^{\rm dense}$ and SFR.
Nevertheless, in spite of the ambiguity in the interpretation of HCN luminosity, $M_{\rm gas}^{\rm dense}$ and SFR are indeed correlated (more in Footenote 3; Jiao et al. submitted).
} in external galaxies. 
This tight linear relation was found to be valid over an extremely wide range of $M^{\rm dense}_{\rm gas}$, which makes it a broadly applied recipe of star formation in the studies of galaxy formation and evolution \citep{Schinnerer2024}.
As unresolved observations on individual galaxies always cover a large number of star-forming molecular clouds, the Gao-Solomon relation may be understood by the central limit theorem (CLT).
For example, one might consider that each unit of $M^{\rm dense}_{\rm gas}$ in the molecular clouds provides an equal chance to randomly sample a universal stellar initial mass function (IMF) through an identical process (hereafter referred to as the stochastic picture).


It became puzzling when resolving a linear correlation between SFR and $M^{\rm dense}_{\rm gas}$ in samples of Galactic molecular gas clumps or clouds  \citep{Wu2005,Wu2010,Lada2012}\footnote{
Note that \citet{Lada2012} re-scaled the ordinary Gao-Solomon relation upward to match the star-formation rated derived based on YSO countings in molecular clouds in the Solar neighborhood.
}.
It implies that the low-mass (e.g., Taurus molecular cloud, $\sim2.0\times10^{4}$ $M_{\odot}$; \citealt{Goldsmith2008ApJ...680..428G}) and high-mass (e.g., W49A,  $\sim2.0\times10^{5}$ $M_{\odot}$; \citealt{Lin2016}) star-forming clouds follow a universal efficiency to convert their $M^{\rm dense}_{\rm gas}$ to the masses of the forming stellar cluster ($m^{\rm cluster}_{\rm star}$), in spite of the very different stellar mass distribution.
In the stochastic picture, the reason why the Taurus molecular cloud is not forming O-type stars is explained by its limited $M^{\rm dense}_{\rm gas}$.
Since the IMF has a power-law profile at the high-mass end (e.g., \citealt{Kroupa2001MNRAS.322..231K}), the low $M^{\rm dense}_{\rm gas}$ leads to a very low probability of obtaining samples from the high-mass end of the IMF.
If it is this case, then observing ten Taurus mass molecular clouds will still provide a chance to detect a number of forming OB stars that are as massive as those in W49A; and the overall SFR of the ten low-mass molecular clouds is the same as the SFR of W49A.
At this moment, the stochastic picture is merely a mathematical implementation of an empirical law, the IMF.
The underlying physics to support this implementation is not yet clear.
Whether or not this picture is realistic is doubtful. 


In this work, we attempt to explain the Gao-Solomon relation with an alternative picture. 
Our main hypothesis is that the mass of the most massive core ($M^{\rm max}_{\rm core}$) in a molecular cloud is determined by the mass of the parent gravitational bound gas reservoir ($M_{\rm gas}^{\rm bound}$).
In addition, we hypothesize that the most massive star forms in the most massive core at an order-unity mass conversion efficiency (i.e., the star-forming efficiency).
Finally, motivated by the studies of \citet{Weidner2013}, \citet{Yan2017A&A...607A.126Y}, and \citet{Yan2023}, we hypothesize that the star-formation process does not stochastically sample the IMF.
Instead, there might be robust correlations between the stellar mass distribution in a cluster, the star formation rate, and $m^{\rm max}_{\rm star}$.
These relations represent various underlying physical processes that are also not yet clear but may be investigated in future theoretical studies. 

With these hypotheses, the SFR can be linked to $M_{\rm gas}^{\rm bound}$ (and thus $M^{\rm dense}_{\rm gas}$) through the correlations of (i) SFR vs. $m^{\rm max}_{\rm star}$, (ii) $m^{\rm max}_{\rm star}$ vs. $M^{\rm max}_{\rm core}$, and (iii) $M^{\rm max}_{\rm core}$ vs. $M_{\rm gas}^{\rm bound}$.
The Gao-Solomon relation can be explained if the combination of these three non-trivial correlations occurs to yield an approximately linear correlation between SFR and $M_{\rm gas}^{\rm bound}$, and if $M_{\rm gas}^{\rm bound}\sim M^{\rm dense}_{\rm gas}$ (Jiao et al. submitted)\footnote{
Here, we are pushing forward to use $M_{\rm gas}^{\rm bound}$ to substitute the concept of $M^{\rm dense}_{\rm gas}$ referred in the Gao-Solomon relation (\citealt{Gao2004}).
This is to amend an issue that the ordinary Gao-Solomon relation vaguely defined how {\it dense} the gas needs to be in order to form stars.
As this field is maturing, $M^{\rm dense}_{\rm gas}$ will naturally be replaced by a more specific quantity that is physically meaningful. 
Nevertheless, these two quantities are not unrelated. 
In many but not all star-forming regions, we found $M_{\rm gas}^{\rm bound}\sim M^{\rm dense}_{\rm gas}$ if $M^{\rm dense}_{\rm gas}$ is measured in the same way as \citet{Gao2004} while $M_{\rm gas}^{\rm bound}$ is measured in the say described in Section \ref{result:M_pgb}
(Jiao et al. submitted; also see the related discussion in \citealt{Bonnell2003,Xu2023SDC335,VS2017}).
With this substitution, our discussion is also not subject to the ambiguity in the interpretation of the HCN luminosity (see Footnote 1).
}.

\begin{figure*}[]
\centering
\includegraphics[width=\linewidth]{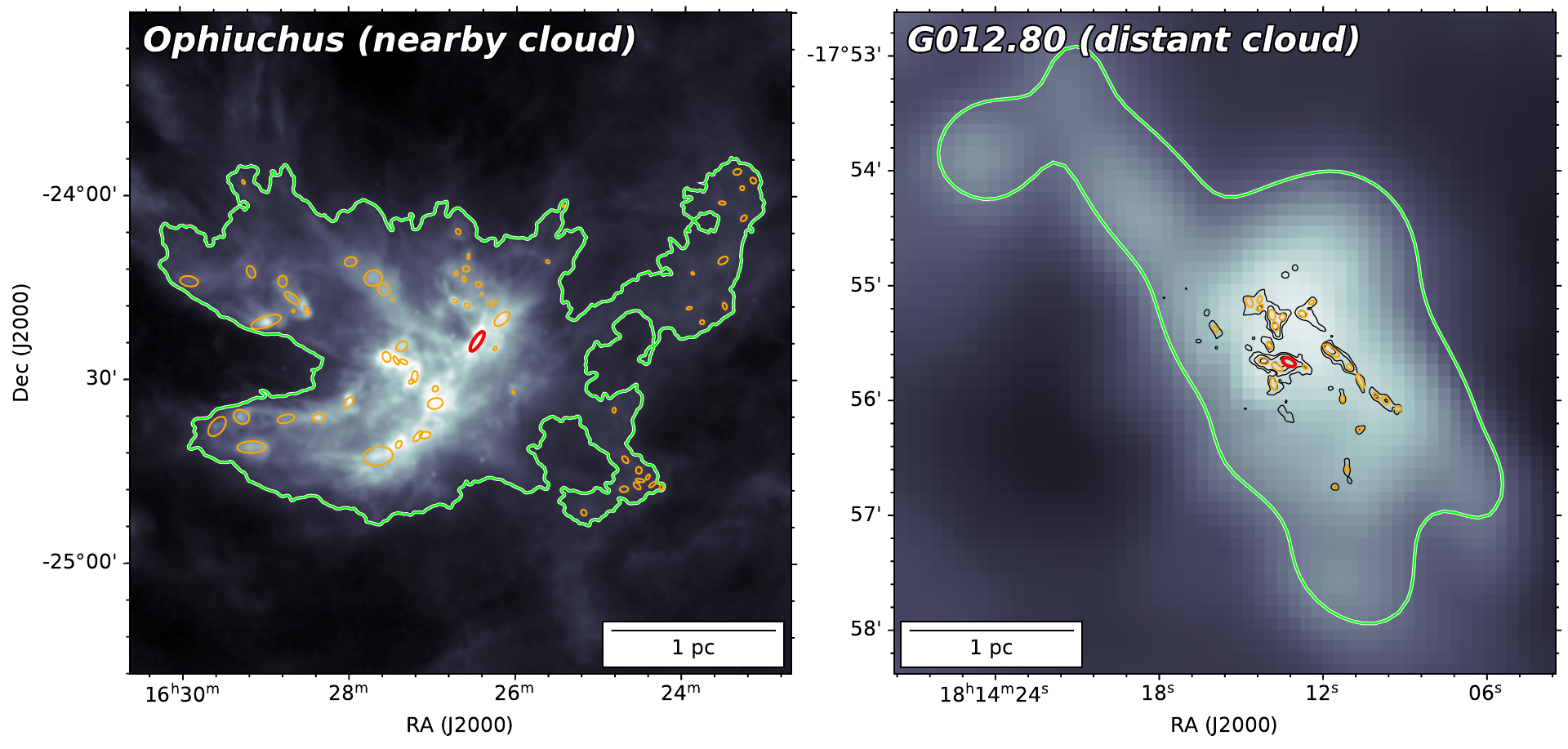}
\caption{The example nearby cloud Ophiuchus (left) and distant cloud G012.80 (right). 
The background shows the H$_2$ column density map in logarithmic scale. The parental gravitational bound gas is defined by the green contours, within which the $M_{\rm gas}^{\rm bound}$ is calculated. The extracted dense cores are marked with yellow ellipses and the most massive one is highlighted in red color. For G012.80, the ALMA-IMF 1.3 mm dust continuum emission is shown in black contours: 0.02, 0.06, and 0.18 \jybeam. 
The scale bar of 1~pc is shown on the bottom. \label{fig:cloudcore}}
\end{figure*}

Our hypothesized scenario was motivated by two previous observational results. 
First, retrieving 139 young star clusters, \citet{Weidner2013} have found that the mass of the most massive star ($m^{\rm max}_{\rm star}$) of a cluster is too tightly correlated with $m^{\rm cluster}_{\rm star}$ to be reconciled with a stochastic sampling of the IMF.
Second, the recent (sub)millimeter dust continuum observations also showed a positive correlation between massive star immediate gas reservoir $M^{\rm max}_{\rm core}$ and parsec-scale molecular clump mass $M_{\rm clump}$ (\citealt{Lin2019,Xu2024QUARKS,Xu2024ASSEMBLE}).

To test our hypotheses, we examined the correlation between $M^{\rm max}_{\rm core}$ and $M_{\rm gas}^{\rm bound}$ based on the {\em Herschel} and Atacama Large Millimeter/submillimeter Array (ALMA) observations on a sample of nearby and distant star-forming molecular clouds.
Then, we discuss how $M^{\rm max}_{\rm core}$ and $M_{\rm gas}^{\rm bound}$ may be related to the SFR, and discuss how these relations might be comprehended by an optimal sampling of the stellar initial mass function (IMF) rather than a stochastic sampling of it (c.f. \citealt{Kroupa2013,Weidner2004,Schulz2015,Yan2017A&A...607A.126Y,Yan2023}).
Finally, we discuss the possibility that the core-formation does not randomly sample the core mass function (CMF) of which the shape is similar with that of the IMF.


The observations are introduced in Section \ref{obs}. 
Our methodology for deriving $M^{\rm max}_{\rm core}$ and $M_{\rm gas}^{\rm bound}$ and their correlation are described in Section \ref{result}.
In Section \ref{discuss:sf_law}, we discuss the links of $M^{\rm max}_{\rm core}$ and $M_{\rm gas}^{\rm bound}$ to the SFR and the Gao-Solomon relation.
A discussion about how the observed $M^{\rm max}_{\rm core}$ may not be consistent with randomly sampling the CMF, is provided in Section \ref{discuss:optimal}.
Our conclusion is given in Section \ref{conclusion}.

\section{Observations}\label{obs}

\subsection{{\it Herschel} dust continuum observation} \label{obs:herschel}

We utilized the archival {\it Herschel} data to calculate the mass of the most massive core formed in a gravitationally bound gas structure and its parental gravitationally bound gas mass. 

For the nearby clouds, we retrieved the {\it Herschel} Gould Belt survey (HGBS) data that were taken at 70/160~$\mu$m using the PACS instrument \citep{Poglitsch2010} and at 250/350/500~$\mu$m using the SPIRE instrument \citep{Griffin2010}.
The HGBS\footnote{Detailed observations and data reductions are available on the HGBS archives webpage: HTTP://gouldbelt-herschel.cea.fr/archives. 
The reduced SPIRE/PACS maps for all nearby molecular clouds also are retrieved from the same website.} took a census in the nearby (0.5 kpc) molecular cloud complexes for an extensive imaging survey of the densest portions of the Gould Belt, down to a 5$\sigma$ column sensitivity of $N_{\rm\ssstyle H_2}$ $\sim$ 10$^{21}$ cm$^{-2}$ or $A_{\mbox{\scriptsize V}}$ $\sim$ 1 \citep{Andre2010}.
All target fields were mapped in two orthogonal scan directions at a scanning speed of 60$''$ s$^{-1}$ in parallel mode, acquiring data simultaneously in five bands.
The angular resolution in this observation mode is 7.6$''$ at 70~$\mu$m, 11.5$''$ at 160~$\mu$m, 18.2 $''$ at 250~$\mu$m, 25.2$''$ at 350~$\mu$m, and 36.9$''$ at 500~$\mu$m.
Data were reduced using {\it Herschel} Interactive Processing Environment (HIPE) version 7.0.

For distant and massive clouds, we retrieved the 70, 160, 250, 350, and 500~$\mu$m images from the {\it Herschel} Infrared Galactic plane Survey \citep[Hi-GAL;][]{Molinari2010PASP}. 
Since we are interested in both compact and extended structures, we adopt extended emission products, which have been corrected for absolute zero points based on images taken by the {\it Planck} space telescope. 

\begin{figure*}
\centering
\includegraphics[width=0.8\linewidth]{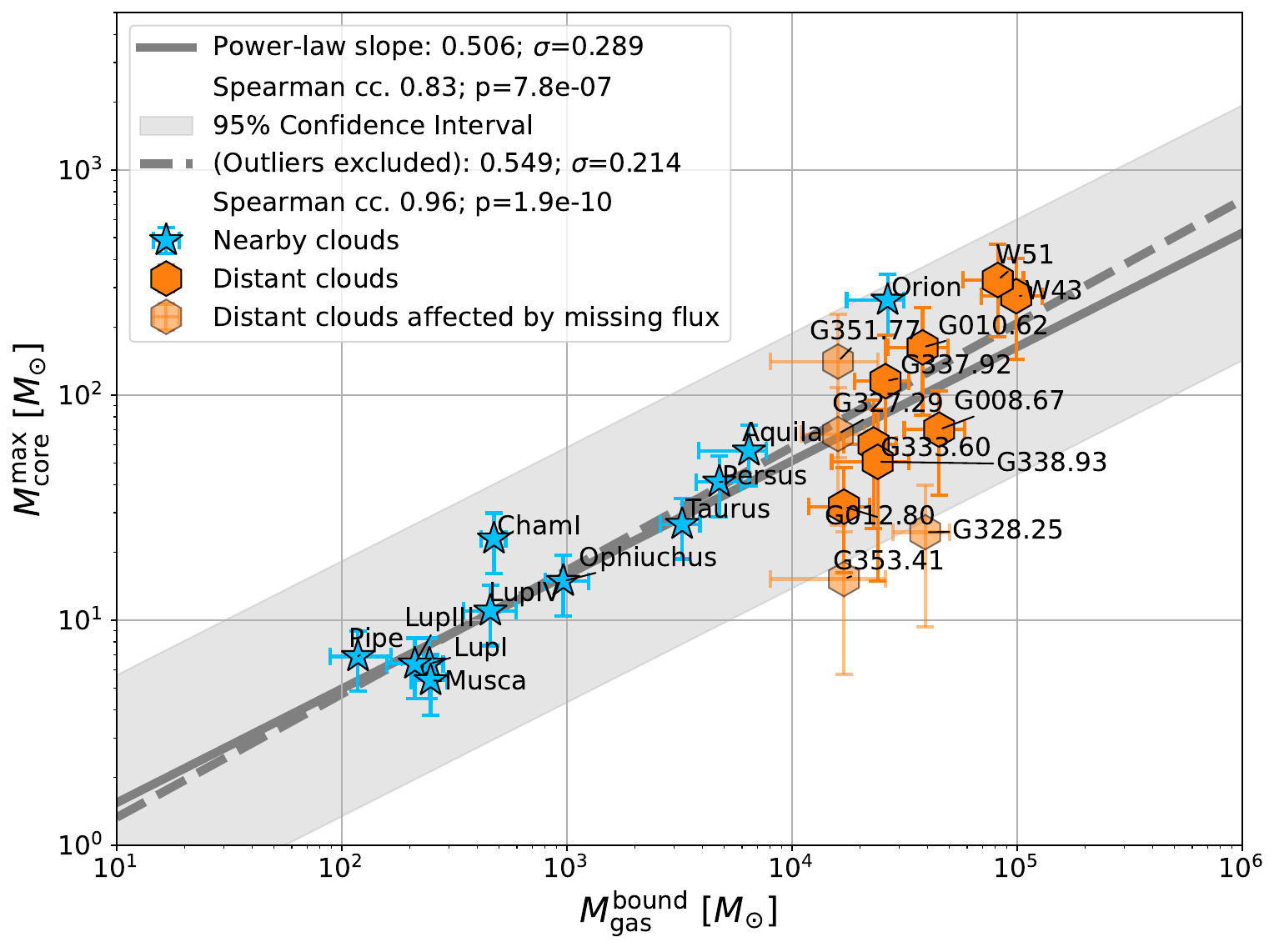}
\caption{The most massive core mass $M^{\rm max}_{\rm core}$ versus its parental gravitational bound gas $M_{\rm gas}^{\rm bound}$. 
The blue stars show the values as well as the uncertainties in the case of nearby clouds, while the orange hexagons stand for massive and distant clouds. 
All the values are listed in Table \ref{tab:mass}.
The black solid line represents the scaling relation for all the cloud sample $M_{\rm gas}^{\rm bound} \propto (M^{\rm max}_{\rm core})^{0.506}$, with a Spearman correlation coefficient of 0.83 and p-value $\ll$ 0.001. 
The standard deviation of linear regression residual $\sigma$ is 0.29 and the gray-shaded region corresponds to the regimes with 95\% confidence interval. 
The $M_{\rm core}^{\rm max}$ measurements of G327.29, G328.25, G351.77, and G353.41 are potentially underestimated due to missing fluxes (Section \ref{result:corr}).
After excluding them, the power-law index becomes a steeper of 0.55 but the correlation becomes more significant with a Spearman correlation coefficient of 0.96 and tighter with a standard deviation of the linear regression residual ($\sigma$) of 0.21.
\label{fig:M-M}}
\end{figure*}

\subsection{ALMA-IMF observation dataset} \label{obs:alma-imf}

To resolve the massive dense cores in distant clouds, we retrieved the ALMA large project ALMA-IMF 1.3~mm continuum data\footnote{\href{https://www.almaimf.com/}{https://www.almaimf.com/}} \citep{Motte2022ALMAIMF-1,Ginsburg2022ALMAIMF-2}. 
Fifteen massive star-forming regions were carefully selected: 1) to be massive enough for sampling the initial mass function; 2) to be located at distances between 2 and 5.5 kpc for both resolving core-like structures and covering the entire extent of the parsec-size clouds with ALMA mosaics; 3) to be the representative protocluster per complex except for two fully covered cloud complexes W51 and W43; 4) to cover various evolutionary stages from IR-quiet to IR-bright \citep{Motte2018ARA&A..56...41M}.

The ALMA-IMF consortium produces the 1.3~mm dust continuum images where the free-free contribution was subtracted based on the H41${\alpha}$ emission (\citealt{GalvanMadrid2024arXiv240707359G}). 
A caveat is that the hydrogen radio recombination lines may have maser effect in some regions (\citealt{MartinPintado1989A&A...215L..13M,JimenezSerra2013ApJ...764L...4J,Zhu2022A&A...665A..94Z}), while we assumed that this does not significantly bias the intensities of free-free emission in the regions we analyzed. 
We used free-free subtracted images released by ALMA-IMF team \footnote{\href{https://dataverse.harvard.edu/dataverse/alma-imf}{https://dataverse.harvard.edu/dataverse/alma-imf}}. 
These images have angular resolution from 0\parcsec3 to 1\parcsec5 to obtained a matched spatial resolution of 2000~AU ($\sim$0.01 pc). The achieved rms noises are 0.03--0.18 \mjybeam, which correspond to a comparable point mass sensitivity of 0.15 \msun~at 1.3 mm. The maximum recoverable scale is up to 0.1--0.2~pc (see Figure 6 in \citealt{Ginsburg2022ALMAIMF-2}).
For most cases, this is adequate for our core-scale studies.
We will discuss the effect of missing flux when it is relevant. 

\section{Results} \label{result}

\subsection{The most massive core} \label{result:M_max}

We apply a commonly-used source extraction algorithm \texttt{SExtractor}\footnote{\href{https://sextractor.readthedocs.io}{https://sextractor.readthedocs.io/}} \citep{Bertin1996SExtractor}.  For each image, we extract sources on the background-subtracted emission maps based on the local threshold set by three times noise rms maps. For more details, please refer to Appendix \ref{app:sep}. As examples, we show the \texttt{SExtractor} extracted cores in Ophiuchus (nearby cloud) and G012.80 (distant cloud) in Figure \ref{fig:cloudcore}. 

For nearby clouds, the $N_{\rm\ssstyle H_2}$ images (derived with SED fitting, Appendix \ref{app:NH2}) were smoothed to a consistent spatial resolution of 0.03 pc. Following the procedure in Appendix \ref{app:sep}, the most massive core mass ($M^{\rm max}_{\rm core}$) can be calculated by summing up $N_{\rm\ssstyle H_2}$ from all the pixels that belongs to the core by
\begin{equation} \label{eq:inte}
    M^{\rm max}_{\rm core} = \mathcal{A} \mu m_{\rm H} \sum^{\in \rm core}_{\rm pixel} N_{\rm\ssstyle H_2},
\end{equation}
where $\mu=2.8$ is the mean molecule weight in the interstellar medium \citep{Kauffmann2008}, $m_{\rm H}$ is the mass of a hydrogen atom, and $\mathcal{A}$ is the area of the pixel. 

For distant clouds, we also smoothed the 1.3 mm free-free emission subtracted continuum images to a consistent spatial resolution of 0.03 pc, i.e., an angular resolution of $\sim$1\parcsec1 to $\sim$3\parcsec1, depending on the distance. 
Following the same core extraction procedure but with the intensity map as input, we obtain the flux density of the most massive cores for the massive cloud sample. 
Assuming a uniform temperature gray body emission from dust cores, the total gas mass can be obtained by
\begin{equation}\label{eq:flux2mass}
\begin{split}
M^{\rm max}_{\rm core} &= - \frac{S^{\rm int}_{\rm 1.3mm} d^2}{\kappa_{\rm 1.3mm}B_{\rm 1.3mm} (T_{\rm dust})} \times \frac{\Omega_{\rm beam} B_{\rm 1.3mm} (T_{\rm dust})}{S^{\rm peak}_{\rm 1.2mm}} \\
&\times \ln\left(1-\frac{S^{\rm peak}_{\rm 1.3mm}}{\Omega_{\rm beam} B_{\rm 1.3mm}(T_{\rm dust})}\right).
\end{split}
\end{equation}
Here, $S^{\rm int}_{\rm 1.3mm}$ is the integrated monochromatic flux of core at 1.3 mm, $d$ is the distance to the cloud, $\kappa_{\rm 1.3mm}=0.01$ cm$^{2}$\,g$^{-1}$ is the specific dust opacity interpolated to 1.3 mm assuming $\beta=1.8$ \citep{Ossenkopf1994}, $B_\nu (T_\mathrm{dust})$ is the Planck function at a given dust temperature $T_\mathrm{dust}$. 
The first term in Equation \ref{eq:flux2mass} represents the optically thin ($\tau_{\rm 1.3mm}\ll1$) dust case. $\Omega_{\rm beam}$ is the solid angle of the beam, $S^{\rm peak}_{\rm 1.3 mm}$ is the monochromatic peak flux. 
We tried three different methods to estimate dust temperature, which are outlined in Appendix \ref{app:temp}.
These three methods yielded largely consistent results in our subsequent statistical analyses.
We adopted the method two as the fiducial one, since the other two methods led to unrealistically large $M^{\rm max}_{\rm core}$ (e.g., $>$10$^{3}$ $M_{\odot}$) for a few sources (Table \ref{tab:method}).
The uncertainty of dust mass comes from specific dust opacity \citep[$\sim36$\%, both from dust opacity and gas-to-dust ratio;][]{Sanhueza2017}, distance ($\sim20$\%) and temperature (Table \ref{tab:method}). The assumed dust temperature and the derived core mass are listed in Table \ref{tab:mass}. The most massive core is highlighted with red color in Figure \ref{fig:cloudcore}. 

We emphasize that in this study, all the core-like structures in nearby and distant molecular clouds were extracted using the same procedure and from maps with the same spatial resolution of 0.03 pc for consistency. 
As shown in Table \ref{tab:mass}, the extracted cores have deconvolved radii from 0.04 to 0.23 pc, with a median value of 0.1 pc. We conventionally refer to these extracted structures as cores hereafter. 
We note that the spatial resolutions we are using are coarser than the native resolutions in the ALMA-IMF and HGBS data.
Therefore, our derived most massive core properties are not necessarily consistent with those found in the previous publications of those consortia that were based on the native resolutions \citep[e.g.,][]{Konyves2015,Louvet2024}.

\subsection{Parental gravitationally bound gas} \label{result:M_pgb}


The column density probability distribution function (N-PDF) is a widely-used diagnosis to quantify the statistical distribution of gas and to identify gravitationally bound structures in clouds \citep[e.g.,][]{Chen2018NPDF,jiao2022RAA}.
Observations and numerical simulations have revealed that the power-law tails of N-PDFs trace the high column density, self-gravitating dense gas that is anticipated to collapse to form stars \citep{kainulainen2013,lombardi2015,schneider2015}. 
This work utilizes N-PDF to assess the self-gravitating mass across a diverse sample of molecular clouds.

By defining the normalized column density on the logarithmic scale,
\begin{equation}
\eta = \ln(N_{H_{2}}/\langle N_{H_{2}} \rangle)
\end{equation}
the distribution can be written as 
\begin{equation}\label{eq:ln+pl}
p_{\eta}(\eta) =\begin{cases}
            M(2\pi\sigma_{\eta}^{2})^{-0.5}e^{-(\eta - \mu)^{2}/(2\sigma_{\eta}^{2})} 
            &(\eta < \eta_{t}) \\
            Mp_{0}e^{-\alpha\eta} &(\eta > \eta_{t})
        \end{cases}  ,
\end{equation}
where $\eta_{t}$ is the transitional column density in logarithmic normalized units.

Considering the sensitivities, map area, as well as the requirement for a last closed contour \citep{Alves2017A&A...606L...2A}, we set an optimal column density cut-off ($N_{\rm cut}$) for each cloud, which is listed in Table \ref{tab:mass}.
We then fit the N-PDF of each cloud above this cutoff using the Maximum Likelihood Estimation \citep[MLE;][]{Clauset2009}.
This approach avoids the need to pre-bin the observed column density distributions, thereby circumventing the potential pitfalls associated with binning \citep{Virkar2012ar}.
To identify the transitional column density (both the logarithmic normalized value, $\eta_{t}$, and the absolute value, $N_{\mbox{\scriptsize threshold}}$) and to estimate the uncertainties of the fitted parameters, we employed a Markov Chain Monte Carlo (MCMC) approach using the Python package {\tt EMCEE} \citep{Foreman-Mackey2013PASP..125..306F}.
The threshold column densities are summarized in Table \ref{tab:mass} and are presented in Figure \ref{fig:cloudcore},  \ref{fig:nmaps_1}, \ref{fig:nmaps_2}, and \ref{fig:nmaps_3}.

We evaluated the mass of the parental gravitationally bound gas, $M_{\rm gas}^{\rm bound}$, by integrating the mass above the column density threshold (the turning point between log-normal and power-law), i.e., 
\begin{equation}
    M_{\rm gas}^{\rm bound} = \int^{+\infty}_{N_{\rm \ssstyle threshold}}M(N)dN,
\end{equation}
which are listed in Table \ref{tab:mass}.

\begin{deluxetable*}{cccc|cccccc}[!ht]
\tablecaption{Parameters of the parent molecular clouds and the most massive cores
\label{tab:mass}}
\tablewidth{2pt}
\linespread{1.2}
\tablehead{
\multicolumn{4}{c}{Parental molecular clouds} & \multicolumn{4}{c}{Most massive cores} & \multicolumn{2}{c}{$M>3$~\msun{} cores} \\
\cmidrule(r){1-4} \cmidrule(r){5-8} \cmidrule(r){9-10}
\colhead{Name} & \colhead{Distance} & \colhead{$N_{\rm cut}$ / $N_{\rm threshold}$} & \colhead{$M_{\rm gas}^{\rm bound}$} & \colhead{RA} & \colhead{Dec} & \colhead{Radius} & \colhead{$M_{\rm core}^{\rm max}$} & $n(M>3$~\msun$)$ & $M_{\rm tot}$ \\
\colhead{} & \colhead{(pc)} & \colhead{($\times10^{21}$cm$^{-2}$)} & \colhead{($\times10^3$\msun)} & \colhead{(J2000)} & \colhead{(J2000)} & \colhead{(pc)} & \colhead{(\msun)} & & \colhead{(\msun)}
}
\startdata
\multicolumn{8}{c}{Nearby clouds} \\
\hline
Aquila & 278$\pm{13}$ & 3.55 / 6.69$^{+0.10}_{-0.04}$ & 6.43$^{+1.25}_{-2.57}$ & 18:29:56.9 & -01:58:29.8 & 0.15 & 56.5$\pm$16.9 & 89 & 673.0 \\
Cham I & 150$\pm{14}$ & 0.63 / 2.94$^{+0.03}_{-0.03}$& 0.48$^{+0.06}_{-0.06}$  & 11:06:31.9 & -77:23:43.5 & 0.18 & 23.0$\pm$6.9 & 8 & 64.2 \\
Lupus I  & 151$\pm{10}$ & 0.28 / 1.10$^{+0.20}_{-0.00}$ & 0.24$^{+0.04}_{-0.00}$  & 15:45:28.4 & -34:24:02.6 & 0.12 & 6.4$\pm$1.9  & 10 & 44.6 \\
Lupus III  & 197$\pm{13}$ & 0.45 / 1.61$^{+0.04}_{-0.04}$ & 0.21$^{+0.05}_{-0.05}$  & 16:08:48.1 & -39:07:26.9 & 0.11 & 6.4$\pm$1.9 & 3 & 16.4 \\
Lupus IV  & 151$\pm{10}$ & 0.50 / 0.75$^{+0.01}_{-0.01}$ & 0.45$^{+0.13}_{-0.10}$  & 16:01:34.4 & -41:52:10.4 & 0.20 & 11.0$\pm$3.3 & 4 & 23.6 \\
Musca  & 160$\pm{13}$ & 0.63 / 1.58$^{+0.02}_{-0.02}$ & 0.25$^{+0.04}_{-0.04}$  & 12:25:33.9 & -71:42:06.1 & 0.19 & 5.4$\pm$1.6 & 1 & 5.4\\
Ophiuchus  & 128$\pm{6}$ & 1.41 / 3.25$^{+0.05}_{-0.07}$ & 0.96$^{+0.35}_{-0.20}$  & 16:26:27.8 & -24:23:51.2 & 0.10 & 14.9$\pm$4.5 & 15 & 115.7 \\
Orion A & 399$\pm{19}$ & 0.71 / 2.05$^{+1.95}_{-0.16}$ & 26.58$^{+4.78}_{-9.13}$ & 05:35:13.5 & -05:23:57.5 & 0.14 & 264.1$\pm$79.2 & 249 & 4439.7 \\
Perseus  & 256$\pm{12}$ & 0.56 / 1.50$^{+0.01}_{-0.01}$ & 4.75$^{+1.18}_{-0.99}$  & 03:28:39.2 & +31:18:40.6 & 0.15 & 41.1$\pm$12.3 & 124 & 903.7 \\
Pipe  & 180$\pm{9}$ & 1.58 / 4.18$^{+0.01}_{-0.01}$ & 0.12$^{+0.04}_{-0.02}$  & 17:11:21.3 & -27:26:17.6 & 0.08 & 6.9$\pm$2.1 & 6 & 29.1\\
Taurus  & 156$\pm{7}$ & 1.05 / 1.62$^{+0.03}_{-0.05}$ & 3.25$^{+0.65}_{-0.65}$  & 04:41:26.5 & +25:23:57.2 & 0.23 & 26.7$\pm$8.0 & 48 & 326.9 \\
\hline
\multicolumn{8}{c}{Distant and massive clouds} \\
\hline
G008.67 & 3400$\pm{300}$ & 7.9 / $13.3^{+0.2}_{-0.2}$ & $45^{+13}_{-13}$ & 18:06:19.1 & -21:37:31.45 & 0.08 & 70.1$\pm$34.2 & \nodata & \nodata \\
G010.62 & 5000$\pm{500}$ & 11.2 / $18.7^{+0.1}_{-0.1}$ & $38^{+11}_{-11}$ & 18:10:28.7 & -19:55:49.60 & 0.12 & 163.8$\pm$81.8 & \nodata & \nodata \\
G012.80 & 2400$\pm{200}$ & 19.9 / $35.6^{+0.3}_{-0.3}$ & $17^{+5}_{-5}$ & 18:14:13.7 & -17:55:43.07 & 0.05 & 32.1$\pm$15.6 & \nodata & \nodata \\
G327.29 & 2500$\pm{500}$ &6.3 / $14.7^{+0.4}_{-0.4}$ & $16^{+5}_{-5}$ & 15:53:07.8 & -54:37:06.31 & 0.04 & 67.1$\pm$40.2 & \nodata & \nodata \\
G328.25 & 2500$\pm{500}$ & 6.3 / $14.4^{+0.3}_{-0.3}$ & $39^{+11}_{-11}$ & 15:57:59.8 & -53:58:00.18 & 0.05 & 24.6$\pm$15.3 & \nodata & \nodata \\
G333.60 & 4200$\pm{700}$ &  15.1 / $22.7^{+0.8}_{-0.6}$& $23^{+6}_{-6}$ & 16:22:09.5 & -50:05:58.17 & 0.08 & 60.2$\pm$34.5 & \nodata & \nodata \\
G337.92 & 2500$\pm{500}$ & 7.1 / $15.1^{+0.4}_{-0.3}$ & $26^{+7}_{-7}$ & 16:41:10.5 & -47:08:03.61 & 0.07 & 115.1$\pm$68.8 & \nodata & \nodata \\
G338.93 & 3900$\pm{1000}$ & 9.9 / $14.3^{+0.4}_{-1.7}$ & $24^{+13}_{-9}$ & 16:40:34.2 & -45:41:36.58 & 0.04 & 50.6$\pm$35.8 & \nodata & \nodata \\
G351.77 & 2000$\pm{700}$ &6.4 / $16.1^{+1.8}_{-1.8}$  & $16^{+8}_{-8}$ & 17:26:42.5 & -36:09:17.64 & 0.05 & 140.4$\pm$86.8 & \nodata & \nodata \\
G353.41 & 2000$\pm{700}$ & 6.5 / $21.1^{+0.6}_{-0.8}$& $17^{+9}_{-9}$ & 17:30:26.1 & -34:41:46.84 & 0.04 & 15.3$\pm$9.5 & \nodata & \nodata \\
W43 & 5500$\pm{400}$ &14.1 / $26.1^{+0.2}_{-0.2}$ & $99^{+29}_{-29}$ & 18:47:47.0 & -01:54:26.56 & 0.10 & 275.0$\pm$131.0 & \nodata & \nodata \\
W51 & 5400$\pm{300}$ & 11.2 / $20.7^{+1.2}_{-2.4}$ & $82^{+25}_{-24}$ & 19:23:43.9 & +14:30:34.83 & 0.11 & 323.8$\pm$143.7 & \nodata & \nodata \\
\enddata
\tablecomments{The properties of nearby clouds and distant massive clouds as well as their most massive cores are listed. 
$N_{\rm cut}$: the cut-off column density of N-PDF.
$N_{\rm threshold}$: the transitional column density of N-PDF between log-normal and power-law. The listed uncertainties are derived from the N-PDF fitting.
$M_{\rm gas}^{\rm bound}$: the parental gravitational bound gas. 
RA, Dec: equatorial coordinates of the most massive core. 
$M_{\rm core}^{\rm max}$: the most massive core mass. $n(M>3$~\msun$)$ and $M_{\rm tot}$ ($M>3$~\msun): the number of cores with mass larger than 3~\msun, and their total mass. These two parameters are only counted for nearby clouds.
The full list of the fitted parameters of the N-PDF are presented and discussed in more detail in Jiao et al. submitted.}
\end{deluxetable*}

\subsection{Mass correlation between most massive core and gravitationally bound gas} \label{result:corr}

As shown in Figure \ref{fig:M-M}, the mass of the most massive core $M^{\rm max}_{\rm core}$ is positively correlated with the mass of the parental gravitationally bound gas $M_{\rm gas}^{\rm bound}$ across three magnitudes of $M_{\rm gas}^{\rm bound}$, from $10^2$ to $10^5$ \msun. 
The nearby cloud samples independently depict this correlation while the distant cloud samples populate more measurements in the range of $M_{\rm gas}^{\rm bound}=$10$^{4}$--10$^{5}$ $M_{\odot}$. 

With a linear regression of all the cloud sample, one can obtain the correlation 
\begin{equation} \label{eq:coremass_to_pbgs}
\log(M^{\rm max}_{\rm core}/M_{\odot}) = 0.506 \log(M_{\rm gas}^{\rm bound}/M_{\odot}) - 0.32,
\end{equation} 
with a Spearman coefficient of 0.83 and a p-value of 7.8 $\times$ 10$^{-7}$ that is much lower than 0.001.

We note that the large data scatter exists for the distant cloud samples, which could be partly attributed to missing flux in the ALMA-IMF images (Section \ref{obs:alma-imf}). 
As mentioned by \citet{Ginsburg2022ALMAIMF-2}, there are four fields, G327.29, G328.25, G351.77, and G353.41 with insufficient ($<5$\%) short baselines to recover scales larger than 0.1~pc. 
We found that by excluding these four sources from our sample, the power-law index becomes 0.55, which is $\sim$ 10\% steeper than that derived from the whole sample; the correlation becomes more robust with the Spearman correlation coefficient of 0.96 and the standard deviation of linear regression residual $\sigma$ becomes even tighter as 0.21. However, missing flux may not yield underestimates of $M^{\rm max}_{\rm core}$ by more than a few times. 
The large data scatter in the distant cloud samples is likely largely intrinsic, for example, due to the more profound or more rapid dynamical evolution of the massive cores than the low-mass cores (e.g., due to accretion, fragmentation, and jet/wind driven mass loss \citealt{Offner2014prpl.conf...53O}).

\section{Discussion}\label{sec:discuss}


\subsection{Connection to the star formation relation}\label{discuss:sf_law}

\begin{figure}
\includegraphics[height=6.5cm]{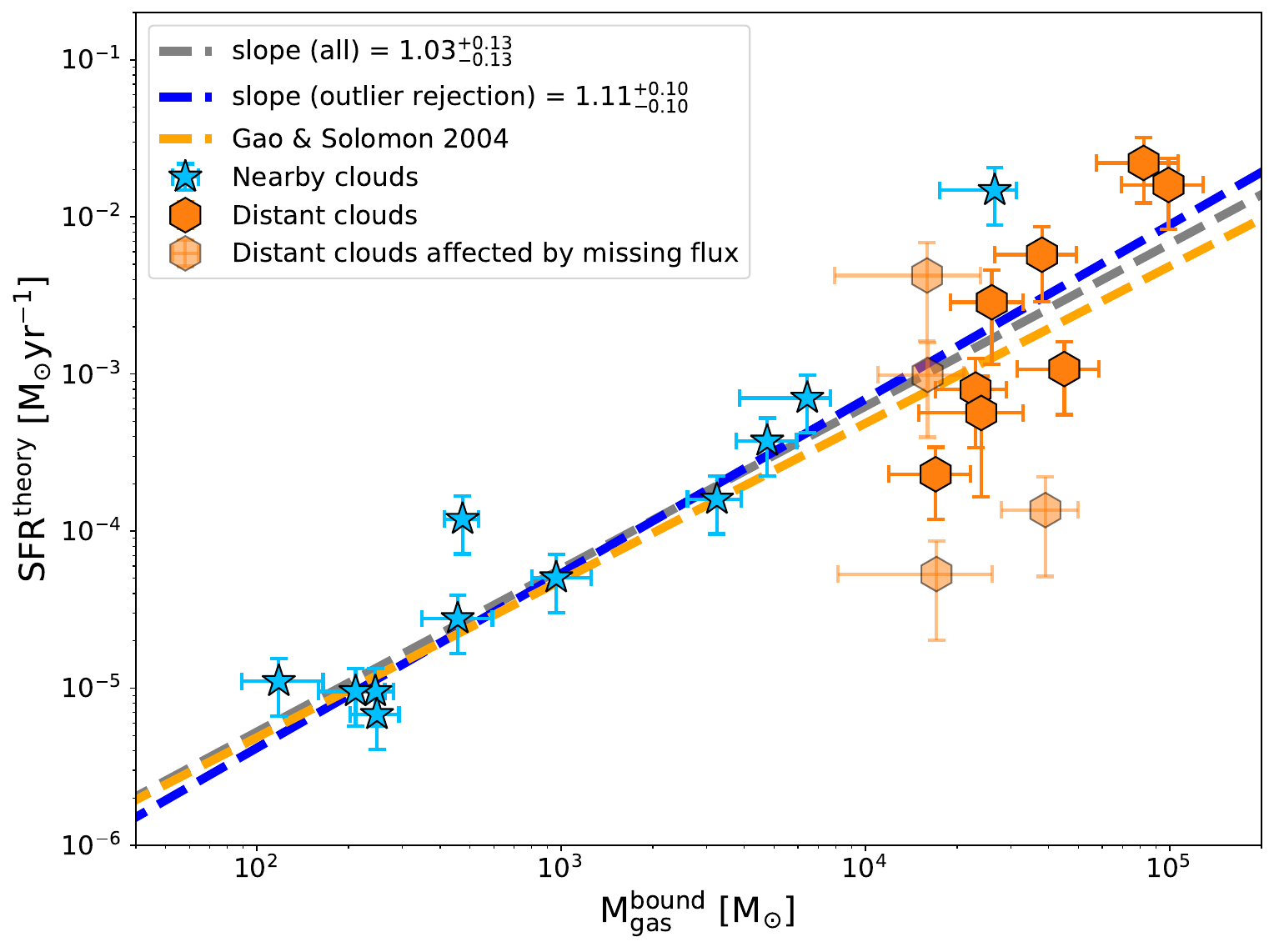}
\caption{
The relation between SFR$^{\rm theory}$ (c.f. Section \ref{discuss:sf_law}) and $M_{\rm gas}^{\rm bound}$ (Section \ref{result:M_pgb}).
The black dashed line shows a linear regression to all data points, while the blue dashed line shows a linear regression to the data points with outlier (e.g., G327.29, G328.25, G351.77, and G353.41) rejection.
The orange line shows the Gao-Solomon relation (\citealt{Gao2004}), which was adjusted upward by a factor of 2.7 as suggested by \citet{Lada2012}.
}\label{fig:sfl}
\end{figure}



If we consider that $M_{\rm gas}^{\rm bound}$ can approximate the $M_{\rm gas}^{\rm dense}$ in the Gao-Solomon (i.e., SFR-$M_{\rm gas}^{\rm dense}$) relation (\citealt{Gao2004}), then a comparison of our observed $M^{\rm max}_{\rm core}$-$M_{\rm gas}^{\rm bound}$ relation (Section \ref{result:corr}) and the Gao-Solomon relation immediately imply that there is a link between $M^{\rm max}_{\rm core}$ and SFR.
Intriguingly, we found that if assuming a constant star-forming efficiency (SFE) of $\sim$30\%, then this link coincides with the theoretical SFR-$m_{\rm star}^{\rm max}$ relation that was derived based on an assumption of optimally sampling the IMF (more below; \citealt{Yan2017A&A...607A.126Y}).

For stars with masses less than 100 $M_{\odot}$, we extracted the theoretical, galactic-scale SFR-$m^{\rm max}_{\rm star}$ relation derived in \cite{Yan2017A&A...607A.126Y} by performing linear regression to the model presented in their Figure 7.
We obtained 
\begin{equation}\label{eq:sfr}
    \log({\rm SFR}/(M_{\odot} {\rm yr}^{-1})) = 2.04 \log(m^{\rm max}_{\rm star}/M_{\odot})-5.80.
\end{equation}
For a star-forming cloud, given the measurement of $M_{\rm gas}^{\rm bound}$, the recipe to theoretically infer the SFR$^{\rm theory}$ then follows:
\begin{enumerate}
    \item Derive $M^{\rm max}_{\rm core}$ based on the observed $M^{\rm max}_{\rm core}$-$M_{\rm gas}^{\rm bound}$ relation (Equation \ref{eq:coremass_to_pbgs}).
    \item Assume $m^{\rm max}_{\rm star}\sim$0.3$\times M^{\rm max}_{\rm core}$
    \item Apply the SFR-$m^{\rm max}_{\rm star}$ relation (Equation \ref{eq:sfr}) mentioned above.
\end{enumerate}
With this recipe, the measurements of $M_{\rm gas}^{\rm bound}$ in our sample then allow us to infer SFR$^{\rm theory}$ (see Appendix \ref{app:sfr} for a comparison with the observed SFR) and a relation between SFR$^{\rm theory}$ and $M_{\rm gas}^{\rm bound}$, which can be compared with the Gao-Solomon relation.
We remark that all relations in this recipe are independent of the Gao-Solomon relation, which is a relation between the observationally constrained SFR and $M_{\rm gas}^{\rm dense}$ (approximately, $M_{\rm gas}^{\rm bound}$).
In this work, we followed \citet{Lada2012} to adjust the ordinary Gao-Solomon relation upward, by multiplying a constant factor 2.7.


As described in Section \ref{intro}, we push forward to use $M_{\rm gas}^{\rm bound}$ to substitute the concept of $M^{\rm dense}_{\rm gas}$ referred in the Gao-Solomon relation. 
We plot the SFR$^{\rm theory}$ and $M_{\rm gas}^{\rm bound}$ of our sample in Figure \ref{fig:sfl}. 
The black dashed line shows a linear regression of all the data points, yielding a slope of 1.03 across all clouds.
The regression line aligns well with the Gao-Solomon relation \citep[][shown as an orange line]{Gao2004}, which indicates that the SFR-$m^{\rm max}_{\rm star}$ relation quoted in our recipe is indeed practical. 
We note that the SFR-$m^{\rm max}_{\rm star}$ relation quoted in our recipe was derived for galactic scale star-formation.
It is intriguing to see that it is also applicable on the spatial scale of molecular clouds. 
We do not yet have a good explanation for it.
This SFR-$m^{\rm max}_{\rm star}$ may be more essential than it was originally thought although it could also be a coincidence.
To clarify it, it requires a deeper understanding (or assumption) about the timescales and efficiency of star formation in a molecular cloud.
At this moment, we can also regard Equation \ref{eq:sfr} as a relation being empirically constrained by the presented data, while the theoretical derivation of \citet{Yan2017A&A...607A.126Y} is one way to comprehend this empirical relation.
The deeper discussion about the underlying physics of the SFR-$m^{\rm max}_{\rm star}$ relation and its applicability is beyond the scope of our present work. 

For more insight, the linear correlation (e.g., SFR $\propto M_{\rm gas}^{\rm bound}$) can be obtained from the combination of $M^{\rm max}_{\rm core}\propto (M_{\rm gas}^{\rm bound})^{0.506}$ (Equation \ref{eq:coremass_to_pbgs}), $M^{\rm max}_{\rm core}\propto$  $m^{\rm max}_{\rm star}$, and SFR $\propto (m^{\rm max}_{\rm star})^{2.04}$ (Equation \ref{eq:sfr}),
\begin{equation}
\begin{split}
{\rm SFR}  &\propto (m^{\rm max}_{\rm star})^{2.04}  \\
&\propto (M^{\rm max}_{\rm core})^{2.04} \\
&\propto ((M_{\rm gas}^{\rm bound})^{0.506})^{2.04} \propto (M_{\rm gas}^{\rm bound})^{1.03}
\end{split}
\end{equation}

This similarity between our deduced SFR$^{\rm theory}$-$M_{\rm gas}^{\rm bound}$ relation and the Gao-Solomon relation supports the notion that star formation, from individual local clouds to entire galaxies, is governed by a similar and straightforward physical principle: the rate at which molecular gas is converted into stars depends on the mass of dense gas within the total gas reservoir \citep[e.g.,][and Jiao et al. submitted]{Lada2012,Evans2014}.
As the SFR-mass diagram is based on the $M^{\rm max}_{\rm core}$-$M_{\rm gas}^{\rm bound}$ relation, the consistency suggests that the linear correlation (a slope of 1) of the Gao-Solomon relation may stem from the highly self-regulated star-forming process when gravity is dominant. 

Finally, we tested the effects of missing flux by rejecting the four potentially affected sources, G327.29, G328.25, G351.77, and G353.41 (Section \ref{result:corr}) from our sample.
In Figure \ref{fig:sfl}, the blue dashed line shows the linear regression to the remaining data points.  
The results shows a slightly steeper slope of $1.11^{+0.10}_{-0.10}$, and the correlation is 30\% tighter.

\begin{figure}
\includegraphics[width=8.67cm]{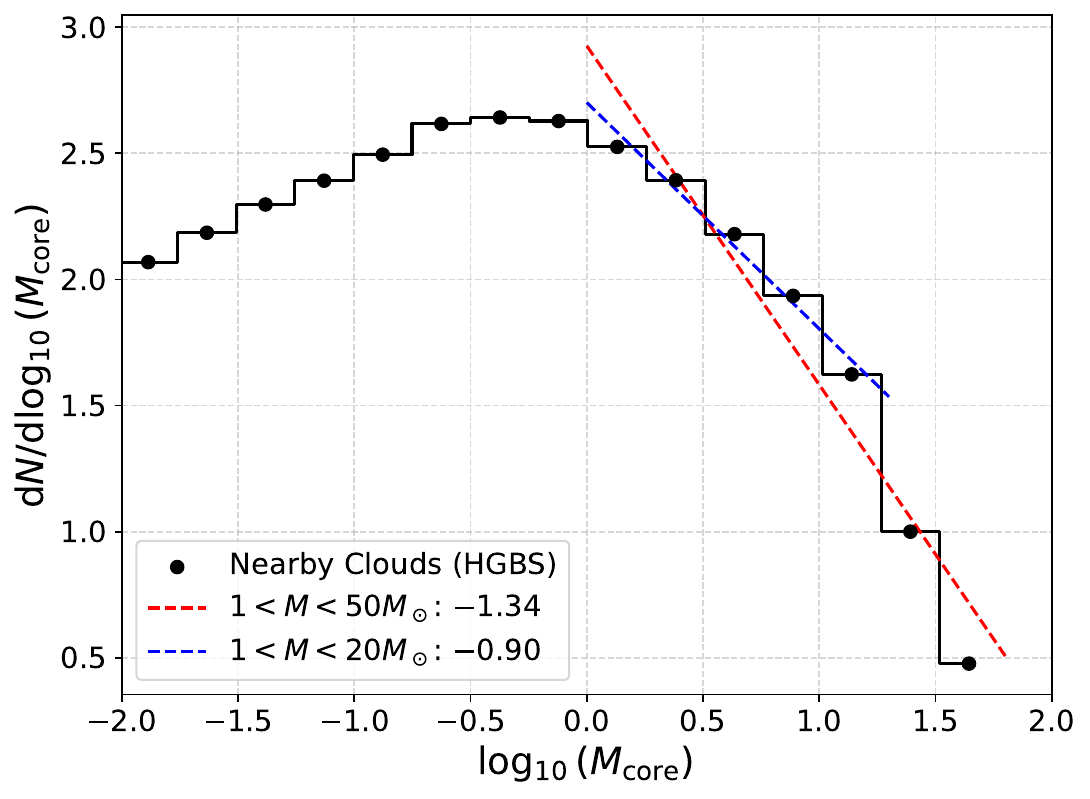}
\caption{CMF for the entire core catalog sample for the nearby clouds. 
Two power-law fittings with mass ranges of $1<M<50 M_\odot$ and $1<M<20 M_{\odot}$ are adopted.}
\label{fig:cmf}
\end{figure}

\subsection{Indication of a non-stochastic sampling of core mass} \label{discuss:optimal}

In the studies of stellar clusters, it was found that the observed correlation between $m_{\rm star}^{\rm max}$ and $m_{\rm star}^{\rm cluster}$ are too tight to be consistent with the mathematical implementation of randomly sampling the IMF \citep{Kroupa2013,Weidner2013}.
As an amendment, an alternative, optimal sampling scheme of the IMF has been recently postulated \citep{Kroupa2013,Weidner2013,Yan2023}.
This new scheme deterministically lists the masses of all stars in a cluster.
In this case, the mass of the most massive star $m_{\rm star}^{\rm max}$ in a stellar cluster directly depends on the total mass of the enclosed stellar cluster $m_{\rm star}^{\rm cluster}$.



Extensive studies have found that the shape of the CMF resembles that of the IMF in both nearby ($<$1 kpc) star-forming regions \citep{Motte1998A&A...336..150M,Alves2007A&A...462L..17A,Konyves2015} and more distant clouds \citep[e.g.][]{Liu2018ApJ...862..105L,Kong2019ApJ...873...31K,Suarez2021ApJ...921...48S,Cheng2024,Louvet2024}.
Specifically, the high-mass end ($M>1M_\odot$) of the CMF can be expressed as a power-law
\begin{equation} \label{eq:cmf}
\frac{d \ N}{d \ {\rm log} M} \propto M^{\Gamma},
\end{equation}
where the power-law index is in the range of [$-$1.35, $-$1.00].
The mass function of the structures we identified (Section \ref{result:M_max}) from the maps of nearby clouds, with the same spatial resolution of 0.03 pc (Figure \ref{fig:cmf}), still appears similar to the canonical IMF/CMF.
In the high-mass end, the power-law indices in various mass ranges are consistent with the range of [$-1.35$, $-1$] seen in other previous studies of IMF/CMF.
Therefore, we think it is appropriate to discuss our identified structures in a context that is similar to how `cores' were discussed in the previous studies.

\begin{figure}
\centering
\includegraphics[width=9.3cm]{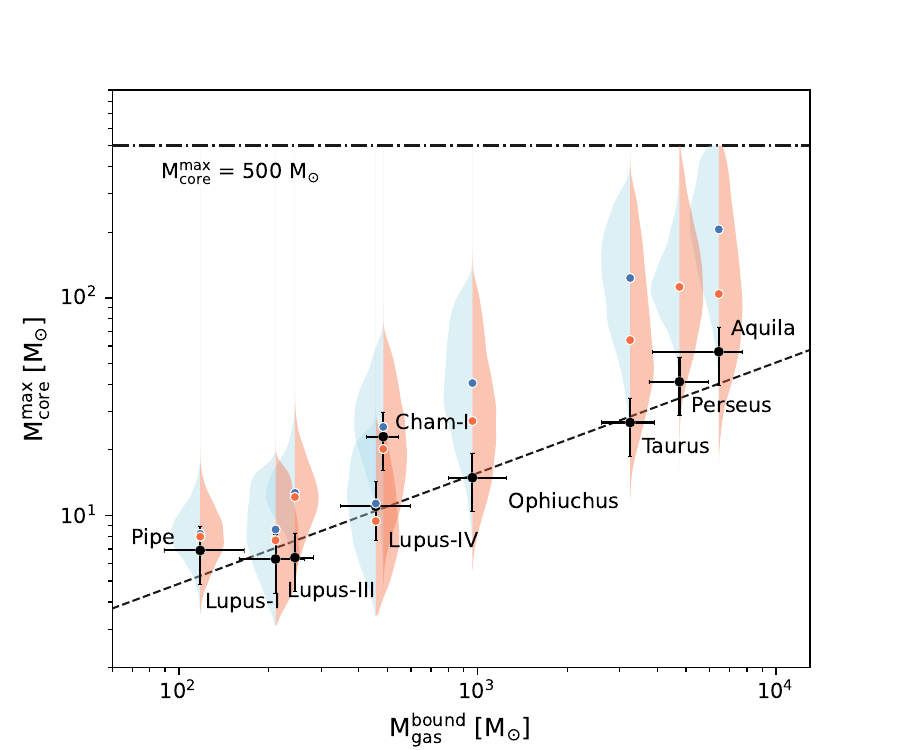}

\vspace{-0.8cm}  

\includegraphics[width=9.3cm]{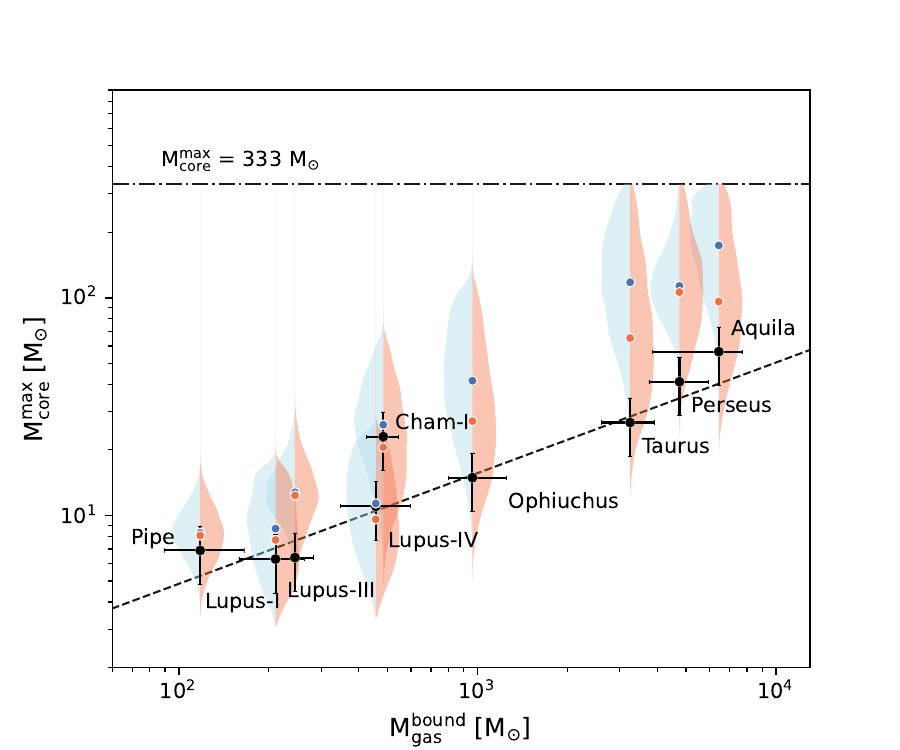}
\caption{
The most massive cores from random sampling (Section \ref{discuss:optimal}).
Upper and lower panels show the cases that the largest possible core masses are artificially limited to $150/{\rm SFE}$ $M_{\odot}$ and $100/{\rm SFE}$ $M_{\odot}$, respectively, where SFE was assume to be 30\%.
The black crosses show the values as well as the uncertainties for the observations. 
The blue and orange violin plots outline the distribution of randomly sampled most massive cores for $\Gamma$ = -1.00 case and $\Gamma$ = -1.35 case, while the median values of $M^{\rm max}_{\rm core}$ from the mock data from random sampling are marked with blue and orange points.
We did not include the Musca cloud in the random sampling because only one core with a mass higher than 3\msun in this region.
}
\label{fig:mc}
\end{figure}

Enlightened by the aforementioned studies on stellar clusters, here, we also discuss whether or not the observed $M_{\rm core}^{\rm max}$ can be consistent with the implementation of randomly sampling the CMF (e.g., Equation \ref{eq:cmf}).
We focus on the nearby clouds observed by the HGBS, in which we obtained a complete sample of $>3M_{\odot}$ cores (given by the $3\sigma$ mass sensitivity). 
For each molecular cloud observed by HGBS, the number of $>3M_{\odot}$ cores is denoted by $n(M>3$~\msun$)$. 
Using the following five steps, we simulated the distribution of $M_{\rm core}^{\rm max}$ assuming that the star-forming processes randomly sample the CMF:
\begin{enumerate}
\item Generate a Core Sample Pool: We generate a sample of cores with mass distribution following the CMF (Equation 8) described by a power-law exponent $\Gamma$ ($\Gamma=-$1.0 and $\Gamma=-$1.35) and a low-mass cutoff as 3 $M_{\odot}$, yielding a sample set of $10^{6}$ cores. 
This initial pool of samples is denoted as set A.
\item Set A is filtered to retain only cores having mass smaller than the total core mass of that cloud and a hypothetical, largest possible core mass. In this work, we tested the two assumptions: (1) the largest possible stellar mass ($m_{\rm up}$) is 150 $M_{\odot}$ (c.f. \citealt{Koen2006MNRAS.365..590K}), and (2) $m_{\rm up}$ is 100 $M_{\odot}$ (c.f. \citealt{Weidner2004}); in both cases, the largest probable core mass is assumed to be $m_{\rm up}/{\rm SFE}$ $M_{\odot}$, where SFE$=$30\%.
This filtered subset is referred to as set B. 
This step ensures that no individual core in the sample exceeds the cloud's total core mass.
\item Random Sampling: From set B, we randomly select a subsample of $n(M>3$~\msun$)$ cores. 
The selected subsample of cores form set C. If the overall core masses in set C does not exceed SFE$\times M_{\rm gas}^{\rm bound}$, we accept set C and proceed to the next step. Otherwise, we redo this step.
\item Identify the Most Massive Core: From set C, the mass of the largest core is registered.
\item Repeat and Analyze: By repeating steps 3 and 4 for 5000 times, we can account for the stochastic nature of the random sampling. 
The resulting distribution of $M_{\rm core}^{\rm max}$ is analyzed using a kernel density estimation (KDE) plot (see Figure \ref{fig:mc}).
\end{enumerate}

For both $\Gamma=-$1.00 and $\Gamma=-$1.35, we found that the simulations yield broad (vertical) distributions of $M^{\rm max}_{\rm core}$ (Figure \ref{fig:mc}).
Based on this result, if the core formation is indeed consistent with a random sampling of the CMF, one would anticipate a large scatter in the observations of $M^{\rm max}_{\rm core}$ versus $M_{\rm gas}^{\rm bound}$; in addition, the degree of scatter is not sensitive to the assumed value of $\Gamma$.
In Figure \ref{fig:mc}, the standard deviation of the observed points from a linear regression line is 0.077.
However, if we replaced the observed $M_{\rm core}^{\rm max}$ values with the those simulated based on randomly sampling the CMF (see above), in the case with a maximum possible core mass $150/{\rm SFE}$ $M_{\odot}$, the mean standard deviation increases to 0.127 for $\Gamma$ $= -$1.00 and 0.139 for $\Gamma=-$1.35; in the case with a maximum possible core mass $100/{\rm SFE}$ $M_{\odot}$, the mean standard deviation becomes 0.121 for $\Gamma = -$1.00 and 0.137 for $\Gamma= -$1.35, respectively.
In other words, the implementation of randomly sampling the CMF yielded a too large scattering of $M_{\rm core}^{\rm max}$ to be consistent with the observed values. 

Moreover, the observed $M^{\rm max}_{\rm core}$-$M_{\rm gas}^{\rm bound}$ relation systematically falls below the expected distribution in the simulations with $\Gamma=-$1.0 and $\Gamma=-$1.35 (Figure \ref{fig:mc}). 
In the following, based on the cases of the Taurus and Perseus regions, two of the nearest known sites of star formation, we elaborate on why the implementation of randomly samping the CMF tend to lead to overestimated $M_{\rm core}^{\rm max}$.
From the HGBS observations, these two regions together have 172 cores that are more massive than 3$M_{\odot}$ while the mass of the most massive core is 41.1 $M_{\odot}$. 
Given the larger number of cores, the $M^{\rm max}_{\rm core}$ from random sampling would significantly exceed the observed $M^{\rm max}_{\rm core}$.
Here, we use a Monte Carlo simulation, using 10$^{6}$ iterations, to test the likelihood of a cloud containing so many cores but no cores exceeding 41.1 $M_{\odot}$. 
Assuming a top-heavy CMF ($\Gamma= -$1.00) at the high-mass end ($M > 3$ $M_{\odot}$), and assuming the maximum possible core mass is $150/{\rm SFE}$ $M_{\odot}$, only 20 out of 10$^{6}$ clouds have no cores above 41.1 $M_{\odot}$; assuming the maximum possible core mass is $100/{\rm SFE}$ $M_{\odot}$, only 21 out of 10$^{6}$ clouds have no cores above 41.1 $M_{\odot}$.
This indicates that the probability for all cores to be less than 41.1 $M_{\odot}$ is $\sim$2 $\times$ 10$^{-5}$.  
Assuming the Salpeter CMF ($\Gamma = -$1.35) case, the probability for all cores to be less than 41.1 $M_{\odot}$ is 7.7 $\times$ 10$^{-3}$ in the case that the maximum possible core mass is $150/{\rm SFE}$ $M_{\odot}$, and is 8.1 $\times$ 10$^{-3}$ in the case that the maximum possible core mass is $100/{\rm SFE}$ $M_{\odot}$.
This test shows that the observed sample of $M^{\rm max}_{\rm core}$ is inconsistent with random sampling from a CMF at a confidence level greater than $>$ 99.1 percent.
Instead, we have to view the core-formation in the Taurus and Perseus regions separately.
Each of them is relatively low-mass clouds that harbor limited numbers of $>$3 $M_{\odot}$ cores and have the correspondingly, specific values of $M^{\rm max}_{\rm core}$.
Extending this argument to observe ten molecular clouds, each as massive as the Taurus molecular cloud, would increase the number of $M >$ 3 $M_{\odot}$ cores tenfold.
However, the $M^{\rm max}_{\rm core}$ would remain constant at the lower mass level, insufficient to form OB stars, challenging the stochastic picture. 

As a summary, observed $M_{\rm core}^{\rm max}$-$M_{\rm gas}^{\rm bound}$ correlation is inconsistent with randomly sampling the CMF in two ways: (1) the observed $M_{\rm core}^{\rm max}$-$M_{\rm gas}^{\rm bound}$ relation is too tight to be consistent with the simulation that randomly sampling the CMF, and (2) the observed $M_{\rm core}^{\rm max}$-$M_{\rm gas}^{\rm bound}$ relation systematically falls below the simulated $M_{\rm core}^{\rm max}$ values.
The number of $M>M_{\rm threshold}$ cores (with $M_{\rm threshold}=3$ $M_{\odot}$ in our case), $n(M>3$~\msun$)$, is uncertain, which can be due to (i) noises (e.g., in the observational data), and (ii) the behavior of the algorithm (i.e., the definition of cores).
Considering random noises in $n(M>3$~\msun$)$ in our simulation will increase the scattering in the simulated $M_{\rm core}^{\rm max}$ values, making them more inconsistent with the observations.
Therefore, it is only necessary to consider the systematic errors in $n(M>3$~\msun$)$.
Suppose that our core-identification algorithm (Section \ref{result:M_max}, Appendix \ref{app:sep}) yielded a large number of spurious identifications in each cloud such that $n(M>3$~\msun$)$ was systematically overestimated, it may indeed explain why the observed $M_{\rm core}^{\rm max}$-$M_{\rm gas}^{\rm bound}$ relation systematically falls below the simulated $M_{\rm core}^{\rm max}$ values.
For instance, in the Taurus molecular cloud, we observe $M_{\rm core}^{\rm max} = 26.7$~\msun while $N = 48$.
To match the simulated median $M_{\rm core}^{\rm max}$ with the observed value through random sampling, the values of $n(M>3$~\msun$)$ need to be 6 for $\Gamma=-$1.0 and 12 for $\Gamma=-$1.35 under the $150/{\rm SFE}$ $M_{\odot}$ core mass limit case, and to be 6 for $\Gamma=-$1.0 and 13 for $\Gamma=-$1.35 respectively under the $100/{\rm SFE}$ $M_{\odot}$ core mass limit case.
The values are significantly smaller than the observed $n(M>3$~\msun$)$, indicating that systematic errors in $n(M>3$~\msun$)$ alone cannot fully explain the vertical offset in Figure \ref{fig:mc}.
In addition, from Figure \ref{fig:mc}, we can also see that it does little help reducing the scattering in the simulated $M_{\rm core}^{\rm max}$. 

\section{Conclusion}\label{conclusion}


Based on the archival {\em Hershel} and ALMA observations, we measured the masses of the most massive cores ($M^{\rm max}_{\rm core}$) and the mass of the gravitational bound gas ($M_{\rm gas}^{\rm bound}$) in the parent molecular clouds from 11 Solar neighborhood clouds and 12 high-mass star-forming regions. 
Our main findings are:

\begin{enumerate}
\item There exists a significant correlation $\log(M^{\rm max}_{\rm core}/M_{\odot}) = 0.506 \log(M_{\rm gas}^{\rm bound}/M_{\odot})-0.32$ for $M_{\rm gas}^{\rm bound}=$10$^{2}$--10$^{5}$ $M_{\odot}$.
\item If we randomly draw samples from a core mass function that has the same slope as the Salpeter IMF ($\Gamma=-$1.35) or a top-heavy IMF ($\Gamma=-$1.00) until there are as many $>3$ $M_{\odot}$ cores as any specific Solar neighbor cloud we analyzed, the $M^{\rm max}_{\rm core}$ in the samples we drew will considerably exceed the $M^{\rm max}_{\rm core}$ observed in that Solar neighbor cloud, and the samples we drew appear too largely scattered to be consistent with the observational data.
\item For the 23 star-forming regions we analyzed, if we assume a universal 30\% efficiency of converting $M^{\rm max}_{\rm core}$ to the mass of the most massive star ($m^{\rm max}_{\rm star}$) and then estimate the star formation rate (SFR)  based on the  $\log$(SFR/($M_{\odot}$ yr$^{-1}$)) $=$ 2.04 $\log$($m^{\rm max}_{\rm star}$/$M_{\odot}$) $-$ 5.80 relation inferred from an optimal sampling of the stellar IMF (\citealt{Kroupa2003ApJ...598.1076K,Yan2017A&A...607A.126Y}), the derived SFR will be consistent with the SFR expected from the Gao-Solomon relation if $M_{\rm gas}^{\rm bound}$ can substitute the dense gas mass ($M^{\rm dense}_{\rm gas}$, e.g., traced by HCN line observations; Jiao et al. submitted). 
\end{enumerate}

Based on points 1 and 2, we hypothesize that cluster-formation is a rather deterministic process. 
There may be robust correlations between $M^{\rm bound}_{\rm gas}$, $M^{\rm max}_{\rm core}$, $M^{\rm max}_{\rm star}$, the overall cluster mass, and SFR. 
The underlying physics is not yet clear.
Intriguingly, we found that the SFR-$M_{\rm gas}^{\rm bound}$ correlation may explain the approximately linear SFR-$M^{\rm dense}_{\rm gas}$ relation in the Gao-Solomon theory. 
On the sub-galactic scales (e.g., $<$1 kpc), the linear Gao-Solomon relation may not be understood by a combination of the central limit theory and proportionality.
Instead, the simple proportionality between SFR and $M^{\rm dense}_{\rm gas}$ may be a consequence of the combination of multiple correlations, each of which may be understood in an analytic/deterministic sense. 
This finding may open a window to break down the long-lasting puzzles on star-formation laws into focused and comprehensive components.


\begin{acknowledgments}
We thank the referee for the useful suggestions. 
We thank Dr. Eda Gjergo for very insightful discussion. 
We thank the high-quality ALMA-IMF data products that have been shared online. 
Herschel is an ESA space observatory with science instruments provided by European-led Principal Investigator consortia and with important participation from NASA.
This paper makes use of the following ALMA data: ADS/JAO.ALMA\#2017.1.01355.L 
ALMA is a partnership of ESO (representing its member states), NSF (USA) and NINS (Japan), together with NRC (Canada), MOST and ASIAA (Taiwan), and KASI (Republic of Korea), in cooperation with the Republic of Chile. The Joint ALMA Observatory is operated by ESO, AUI/NRAO and NAOJ. 

S.J. is supported by NSFC grant nos. 11988101 and 12041302, by the National Key R\&D Program of China No. 2023YFA1608004.
F.W.X is supported by the National Science Foundation of China (12033005) and funding from the European Union’s Horizon 2020 research and innovation programme under grant agreement No 101004719 (ORP). 
H.B.L. is supported by the National Science and Technology Council (NSTC) of Taiwan (Grant Nos. 111-2112-M-110-022-MY3, 113-2112-M-110-022-MY3). 
We acknowledge National Natural Science Foundation of China (NSFC) under grants 12173016, 12041305. We acknowledge the Program for Innovative Talents, Entrepreneur in Jiangsu. We acknowledge the science research grants from the China Manned Space Project, CMS-CSST-2021-A08 and CMS-CSST-2021-A07.
\end{acknowledgments}

\facility{Herschel, Planck, JCMT, ALMA}
\software{CASA, Numpy, APLpy}

\appendix

\section{Column density map}\label{app:NH2}

We perform pixelwise single-component modified black-body spectral energy distribution (SED) fitting of the input {\it Herschel} images at five bands. 
All the images were convolved to a common angular resolution of the largest telescope beam (36\parcsec9) and then were regridded to have the same pixel size (4\parcsec0).
For the modified black-body assumption, the flux density $S_{\nu}$ at a certain observing frequency $\nu$ is given by
\begin{equation}
S_{\nu} = \Omega_{m}B_{\nu}(T_{\mbox{\scriptsize d}})(1-e^{-\tau_{\nu}}),
\end{equation}
where $\Omega_{m}$ is the solid angle, $B_{\nu}(T_{\mbox{\scriptsize d}})$ is the Planck function at dust temperature $T_{\mbox{\scriptsize d}}$, and dust opacity $\tau_\nu$ is given by
\begin{equation}
\tau = \kappa_{\nu} \mu m_{\rm\ssstyle H} N_{\rm\ssstyle H_2} / \mathcal{R}
\end{equation}
where $\kappa_{\nu} = 1.0 (\nu/230\,\rm GHz)^{\beta}$ cm$^2$\ g$^{-1}$ is the dust opacity \citep{Ossenkopf1994},
$\mu = 2.8$ is the mean molecular weight, $m_{\rm\ssstyle H}$ is the mass of a hydrogen atom, and $N_{\rm\ssstyle H_2}$ is the molecular hydrogen column density. 
We adopt a dust emissivity index of $\beta$ = 1.8 and a gas-to-dust mass ratio ($\mathcal{R}$) of 100.
The effect of scattering opacity \citep[cf.][]{Liu2019} can be safely ignored in our case, given that we focus on structures on the scale $>$10$^{3}$ au scale, where the averaged maximum grain size is expected to be well below 100 $\mu$m \citep{Wong2016}.

\begin{figure*}[!ht]
\centering
\hspace{-0.3cm}\includegraphics[scale=0.3]{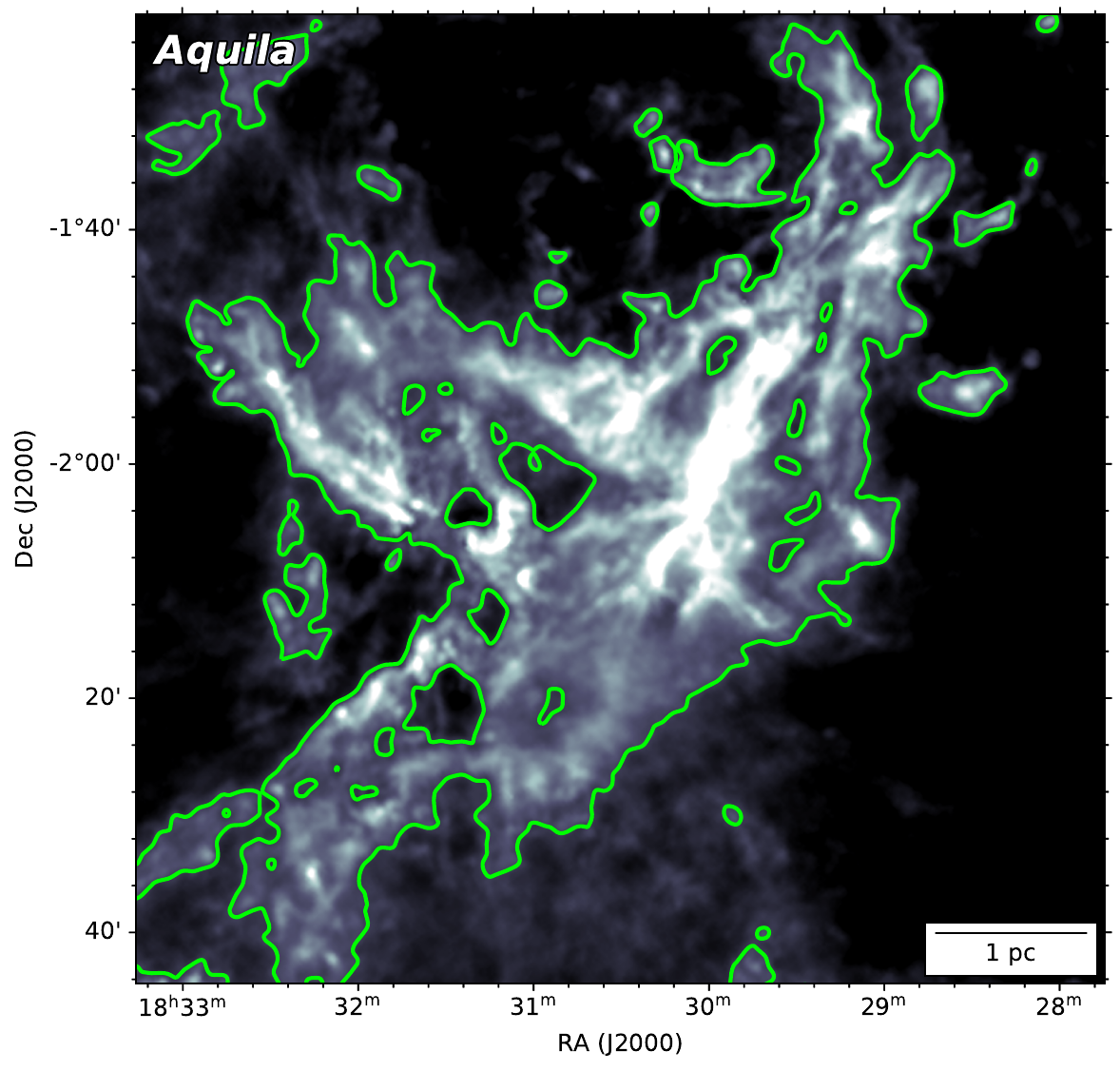} 
\hspace{-0.1cm}\includegraphics[scale=0.3]{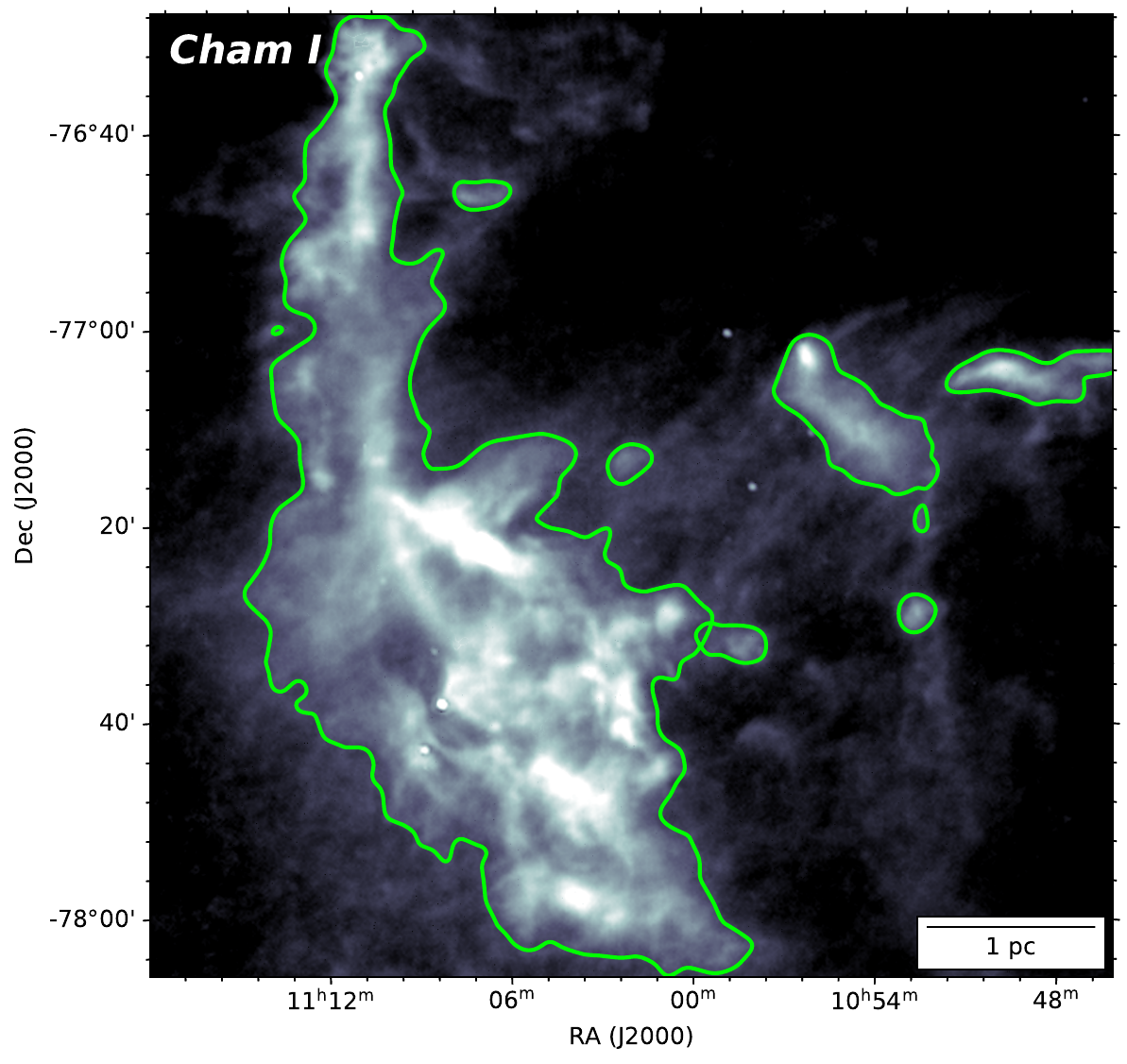} 
\hspace{-0.1cm}\includegraphics[scale=0.3]{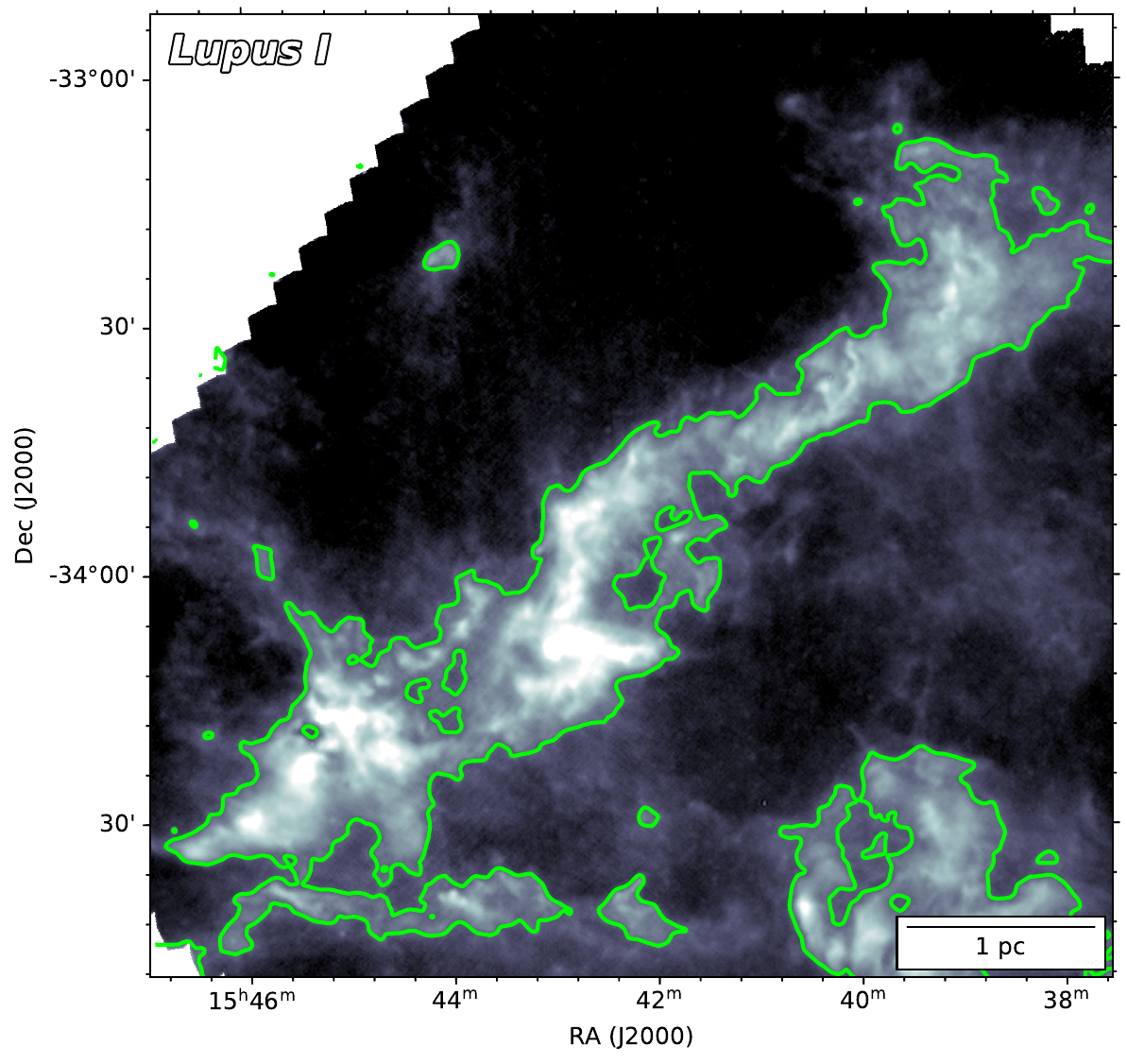}\\
\hspace{-0.3cm}\includegraphics[scale=0.3]{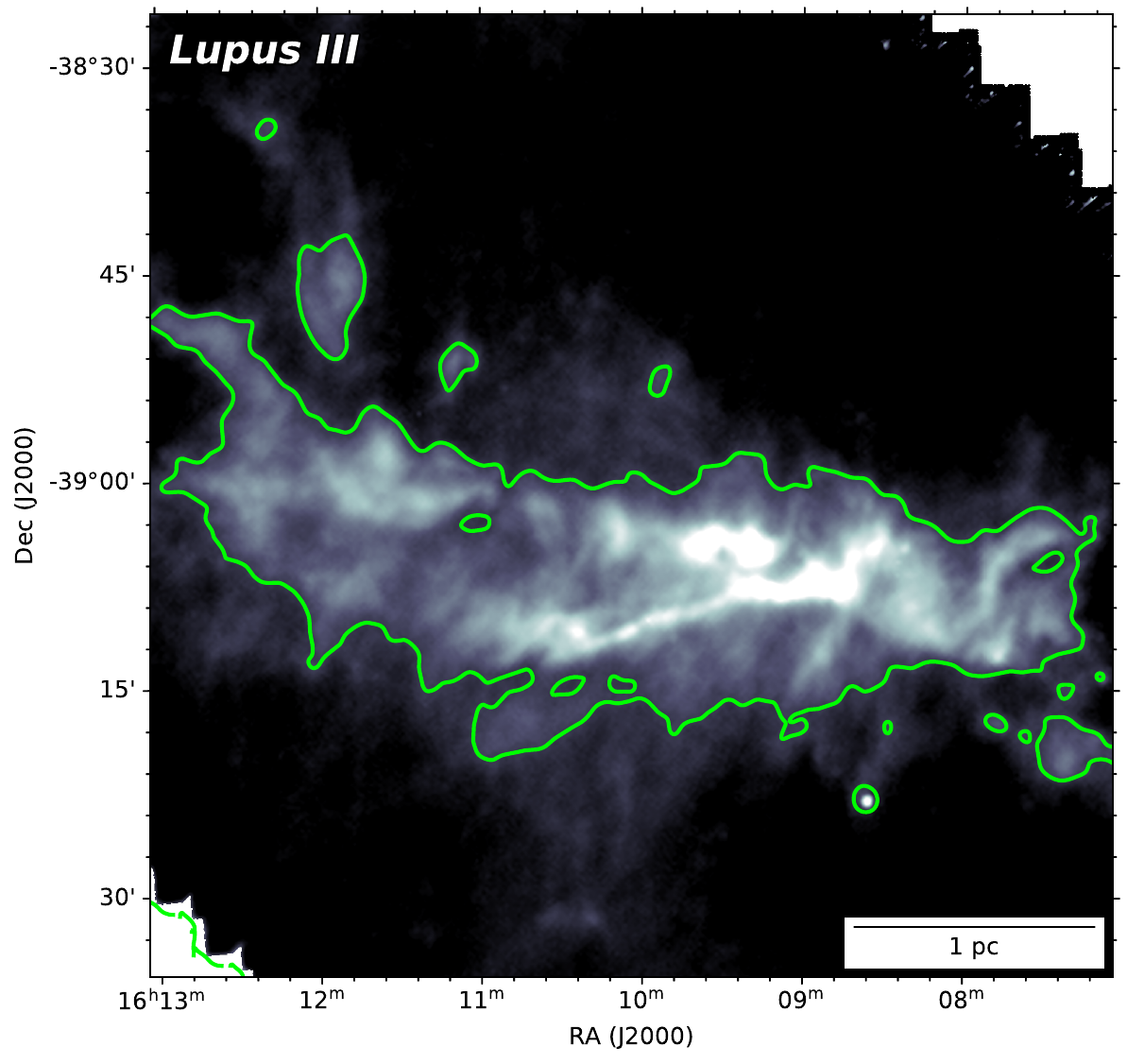} 
\hspace{-0.1cm}\includegraphics[scale=0.3]{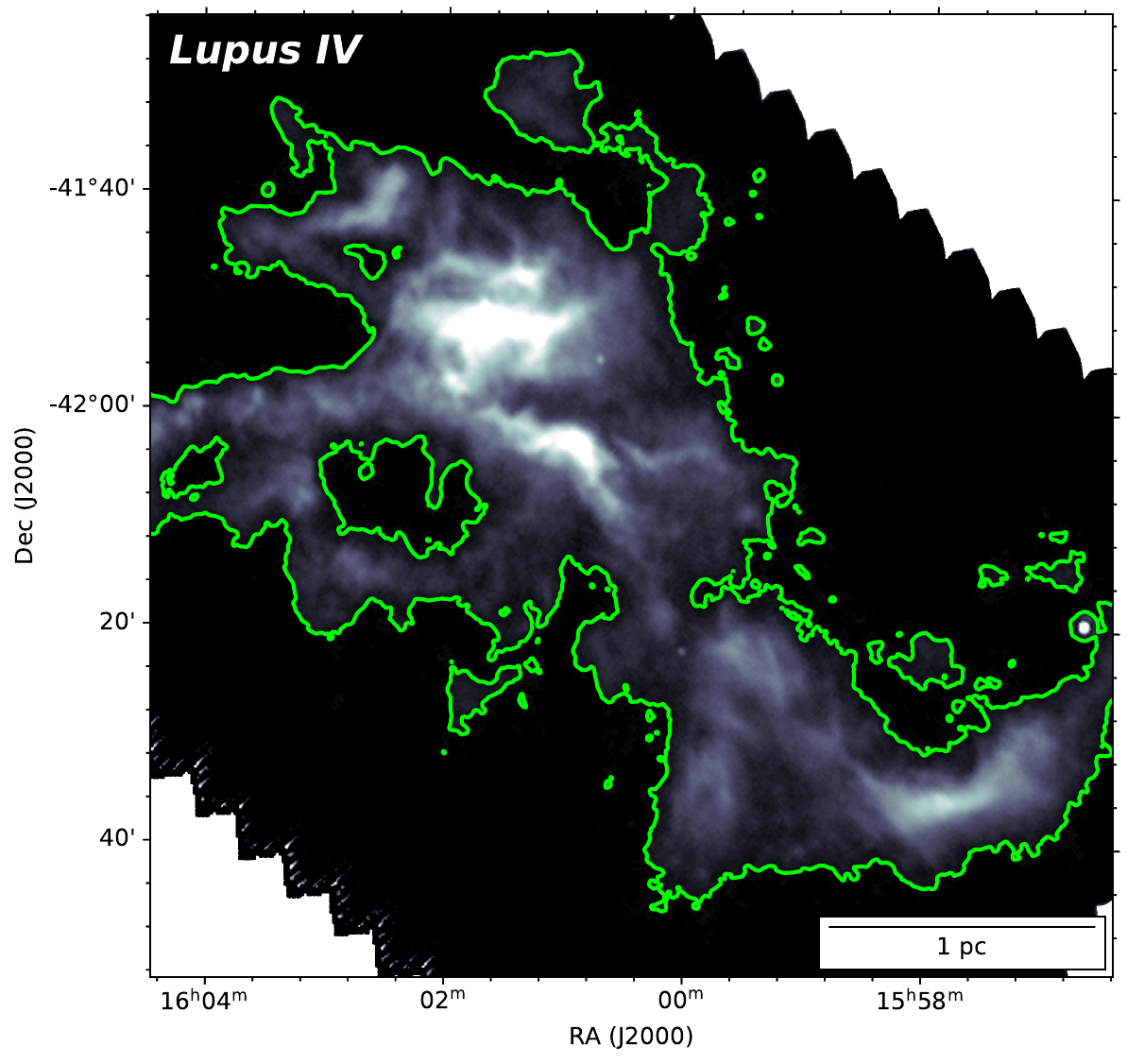} 
\hspace{-0.1cm}\includegraphics[scale=0.3]{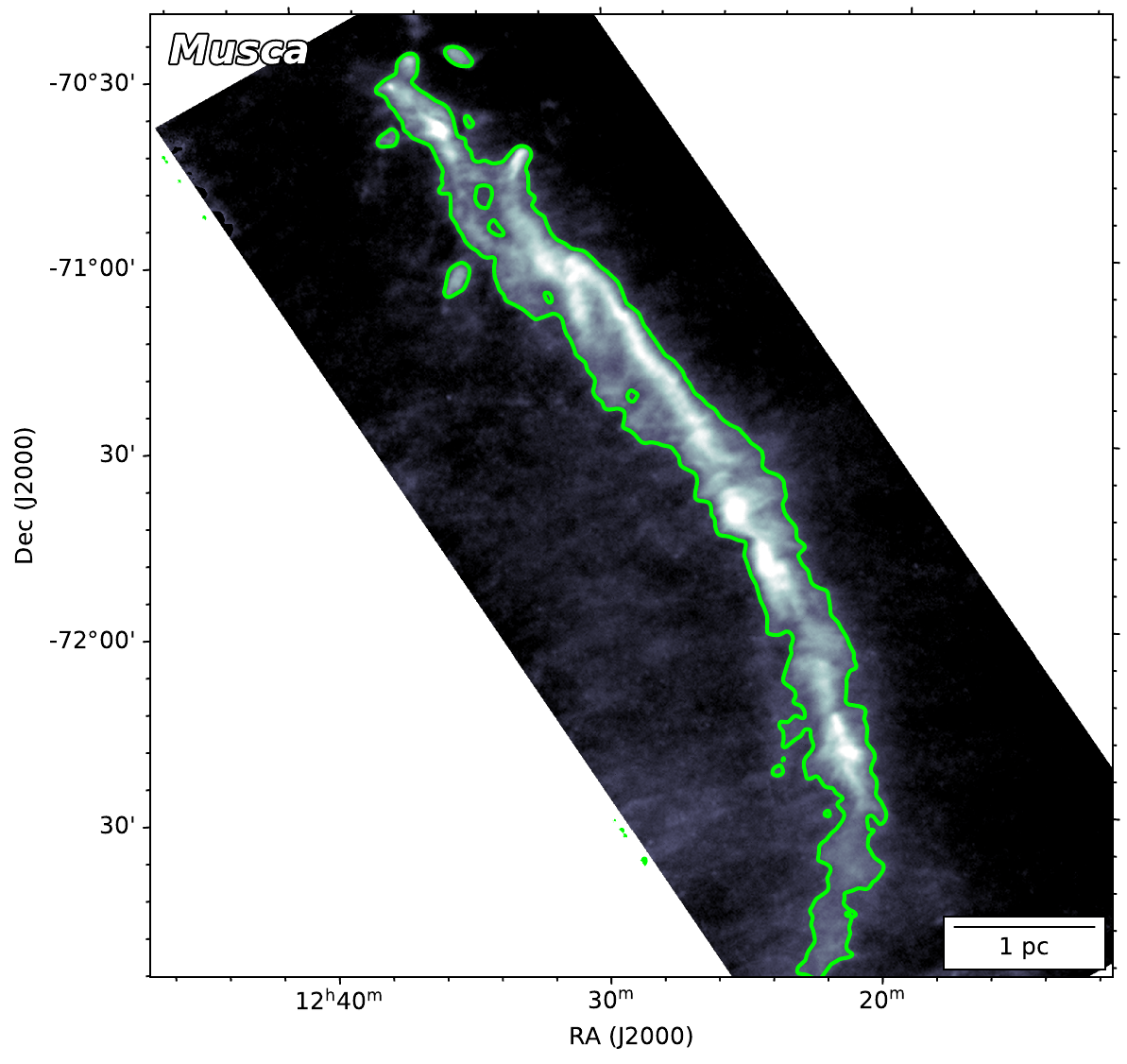}\\
\hspace{-0.3cm}\includegraphics[scale=0.3]{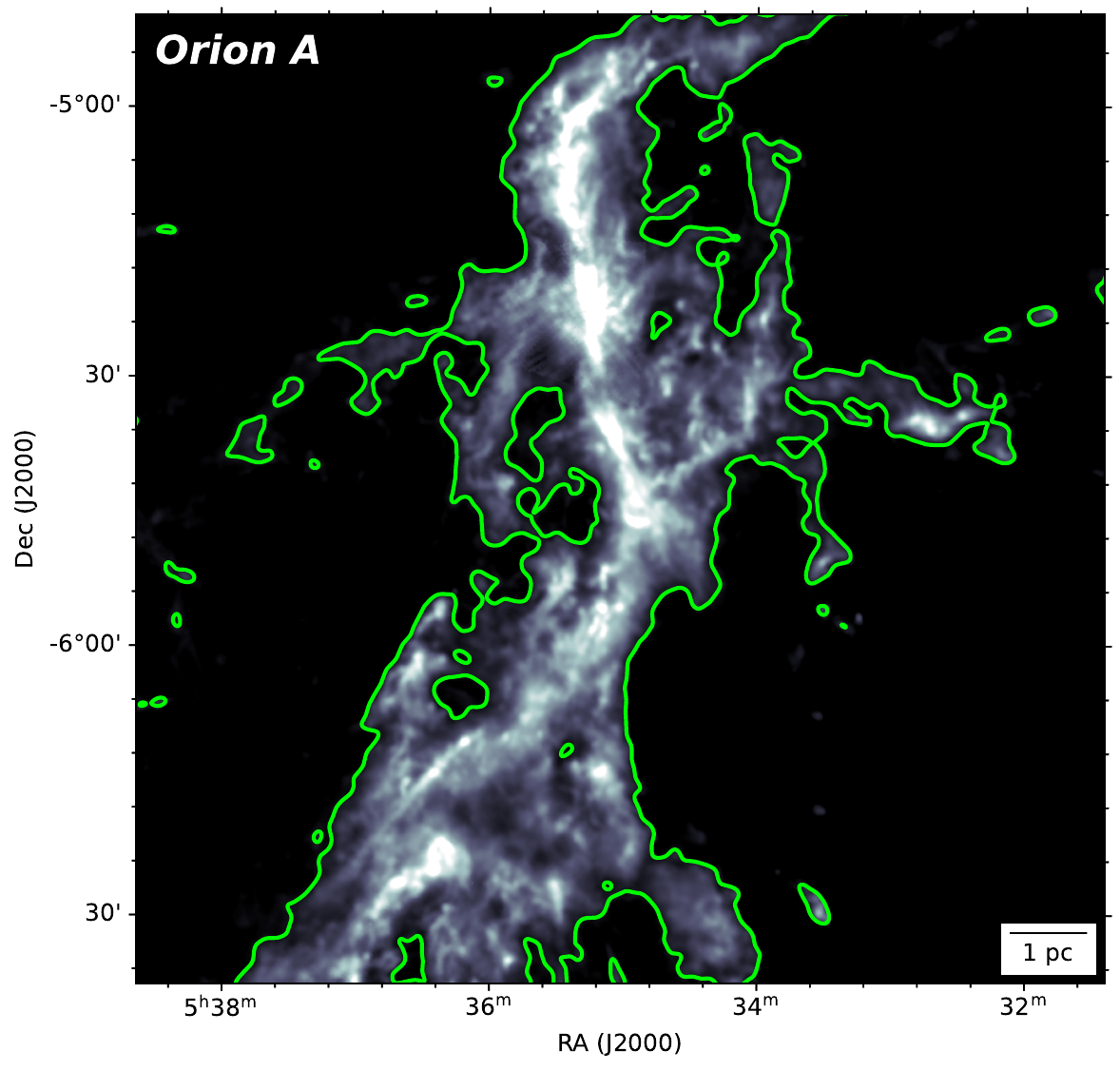} 
\hspace{-0.1cm}\includegraphics[scale=0.3]{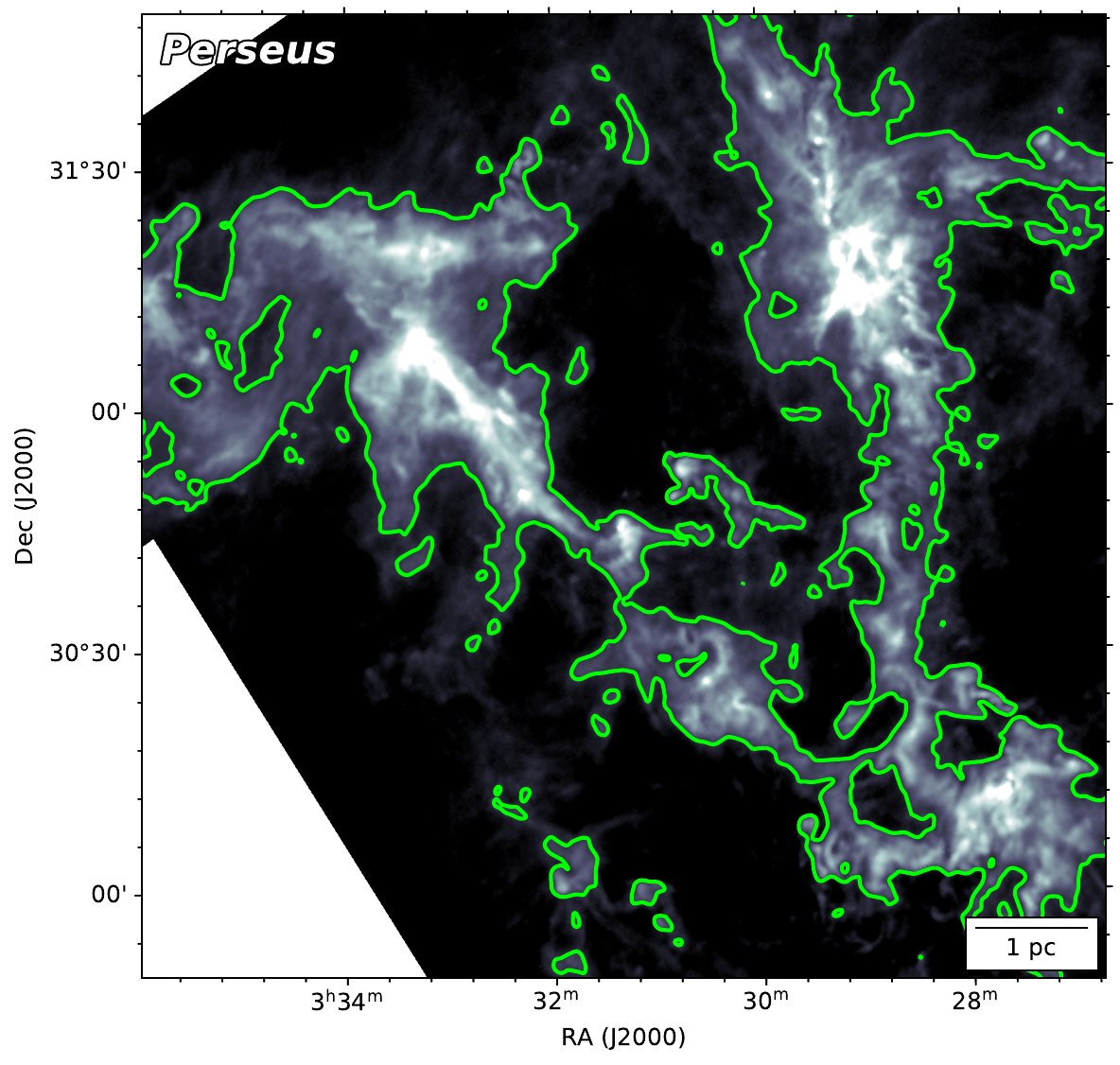} 
\hspace{-0.1cm}\includegraphics[scale=0.3]{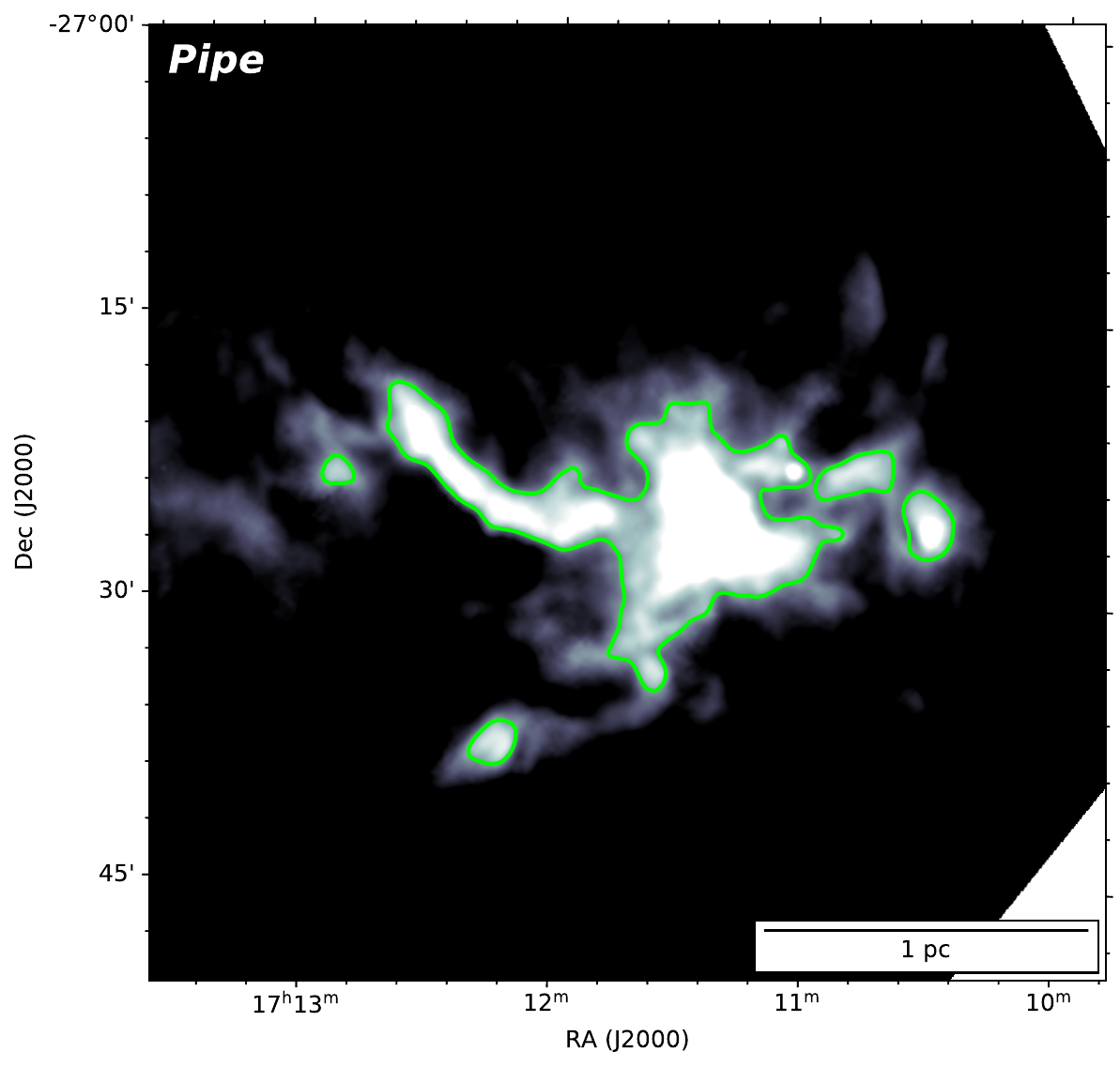}\\
\caption{
The threshold column density (lime contours) and the area associated with $M_{\rm gas}^{\rm bound}$ for each cloud.
}
\label{fig:nmaps_1}
\end{figure*}

\begin{figure*}[!ht]
\centering
\hspace{-0.3cm}\includegraphics[scale=0.3]{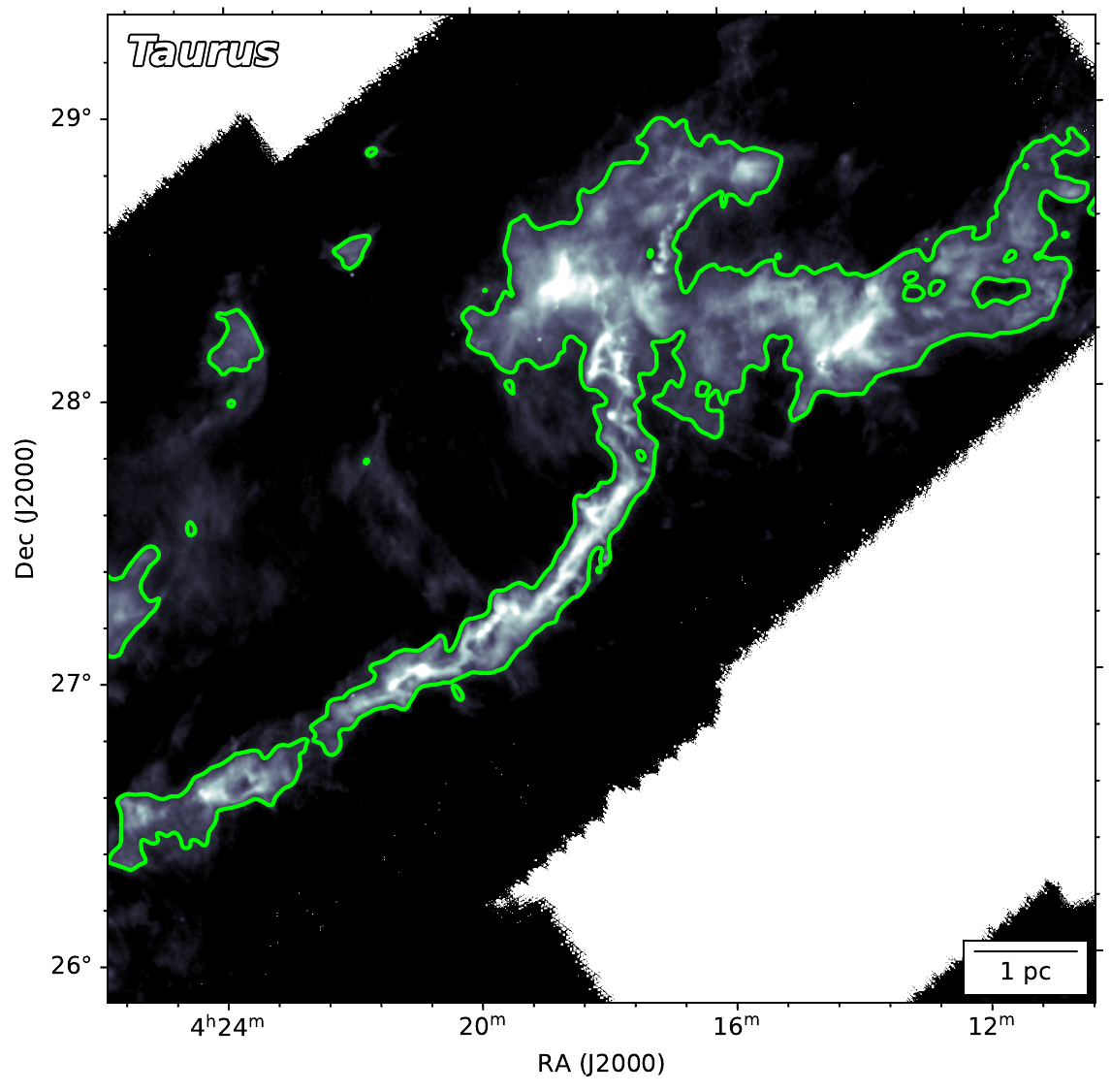} 
\hspace{-0.1cm}\includegraphics[scale=0.3]{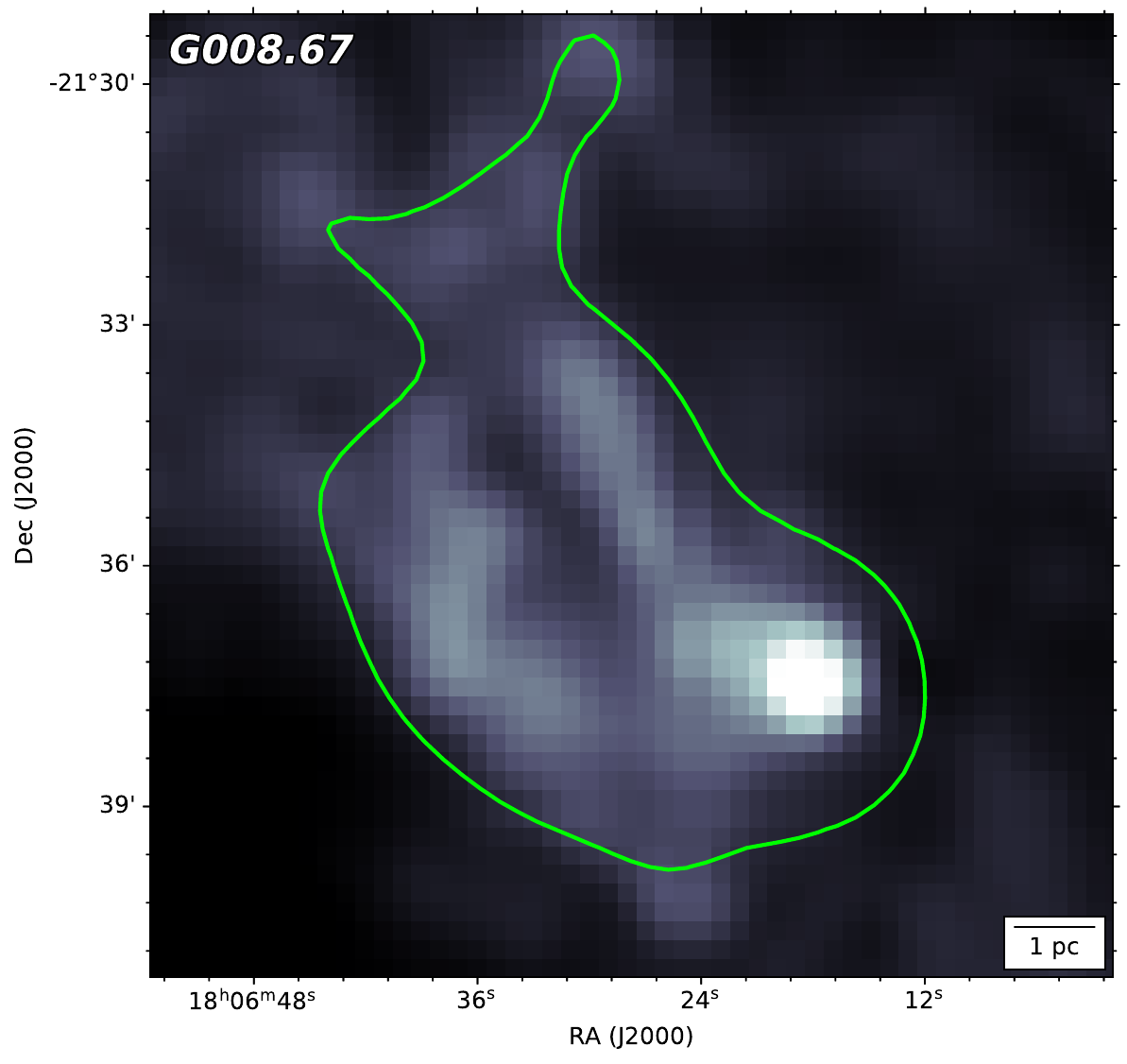} 
\hspace{-0.1cm}\includegraphics[scale=0.3]{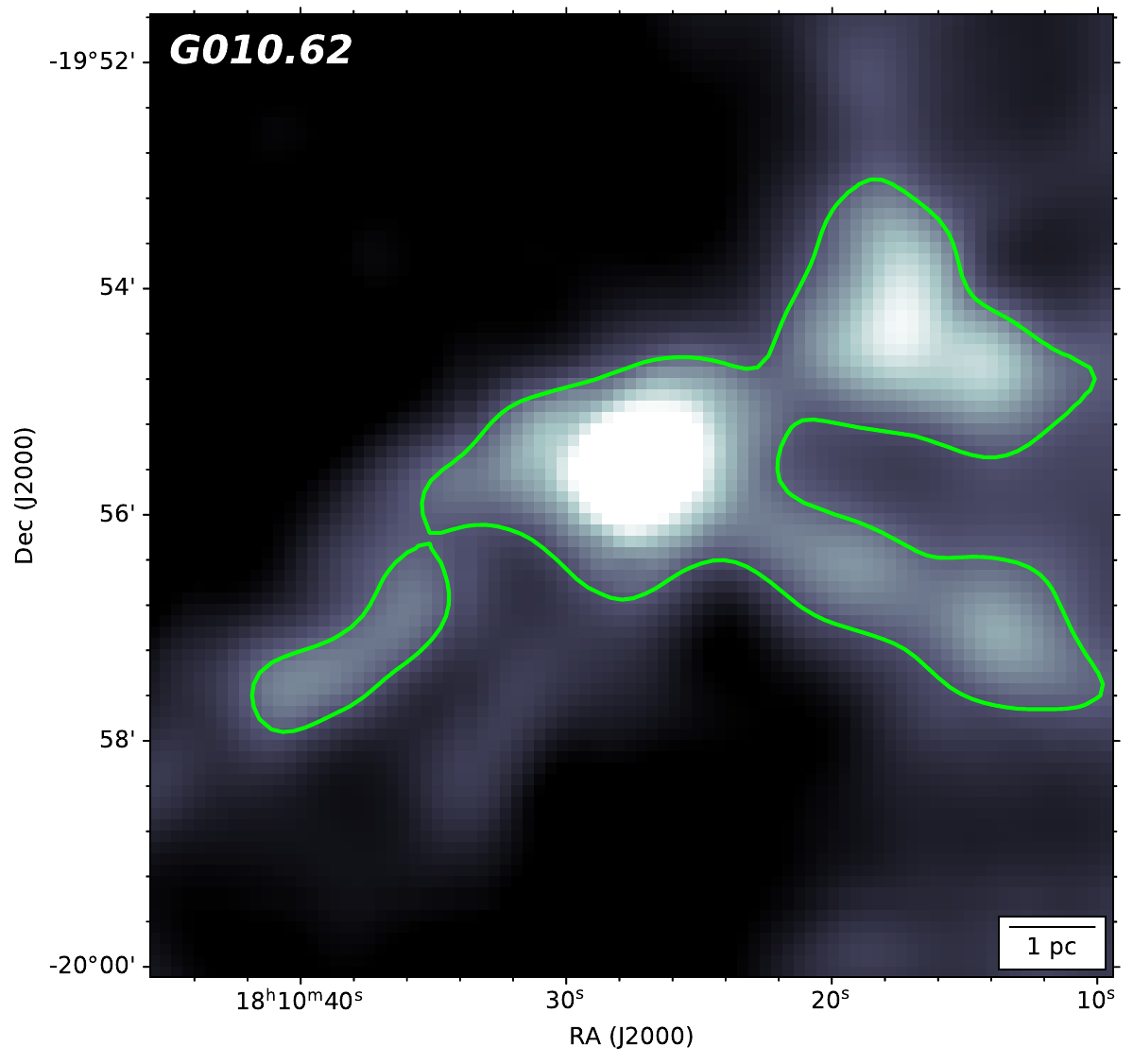}\\
\hspace{-0.3cm}\includegraphics[scale=0.3]{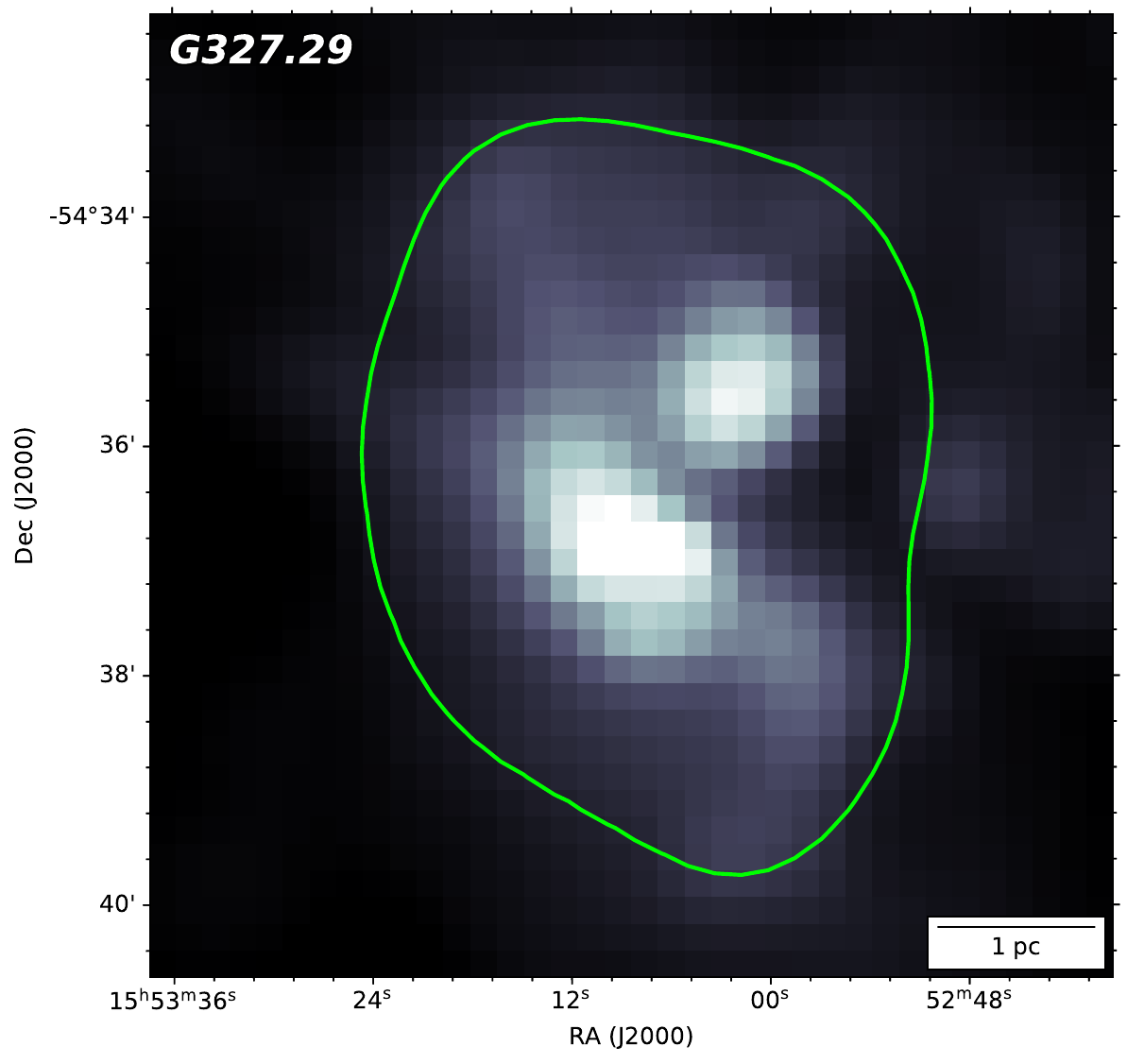} 
\hspace{-0.1cm}\includegraphics[scale=0.3]{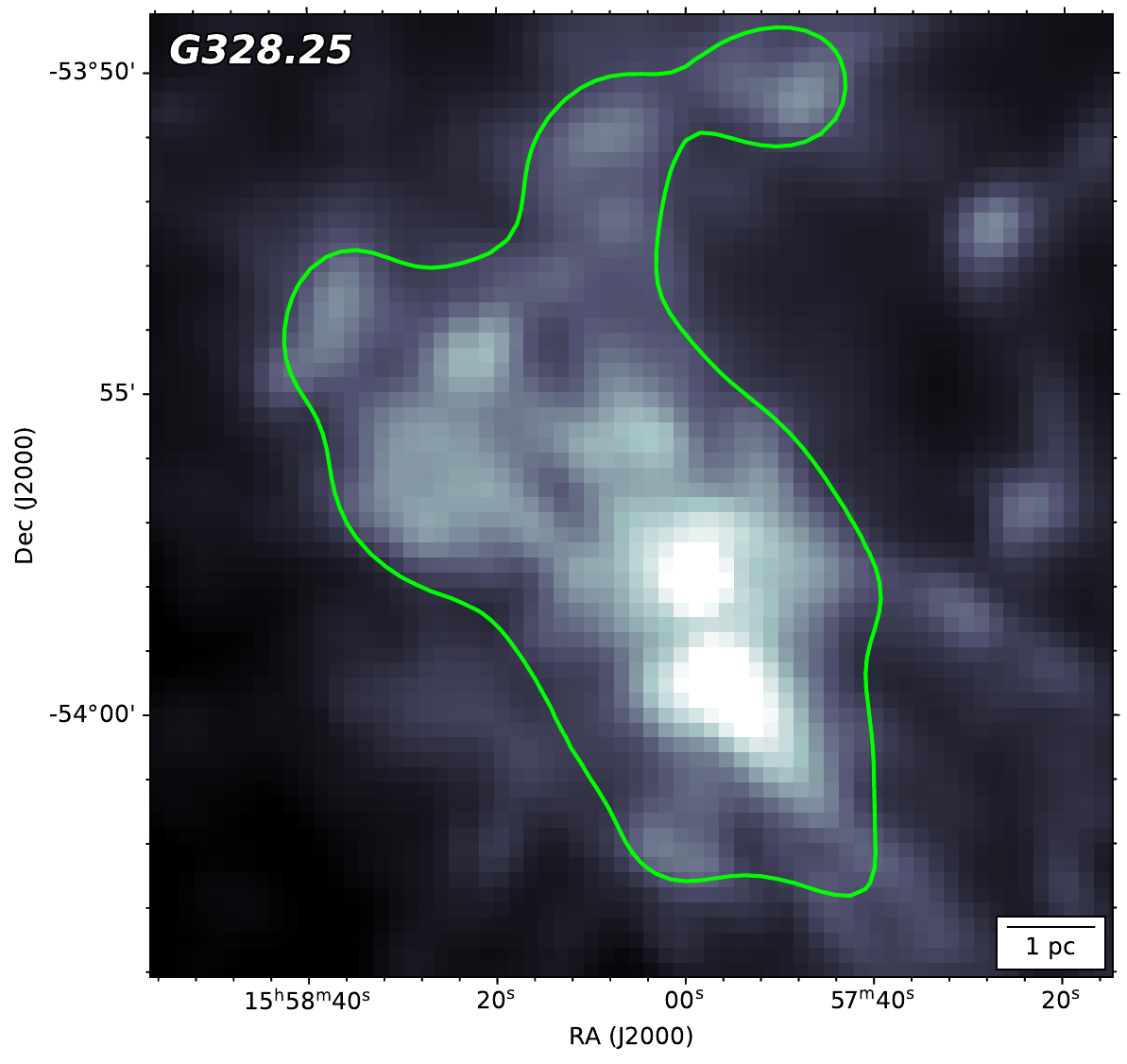} 
\hspace{-0.1cm}\includegraphics[scale=0.3]{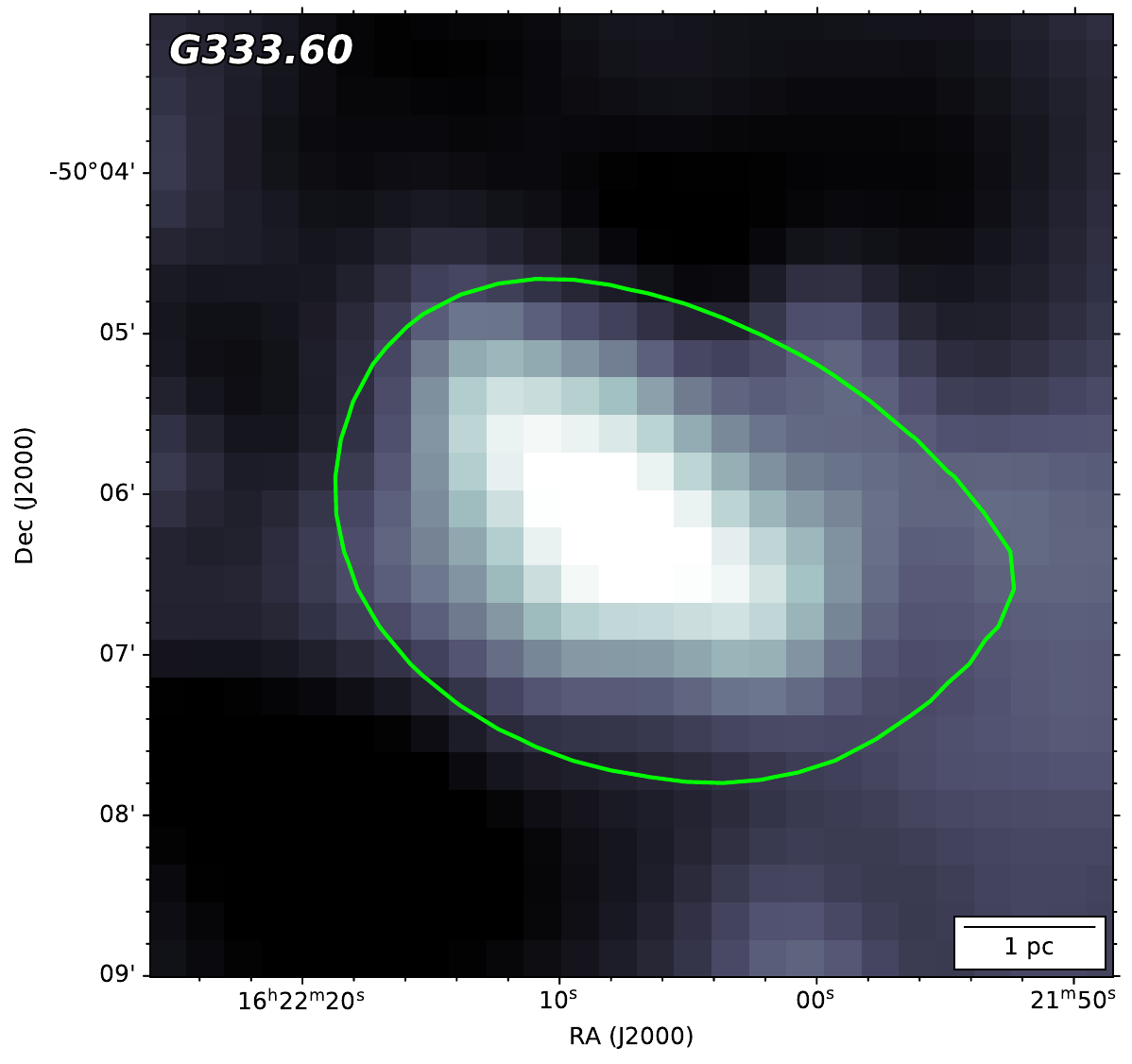}\\
\hspace{-0.3cm}\includegraphics[scale=0.3]{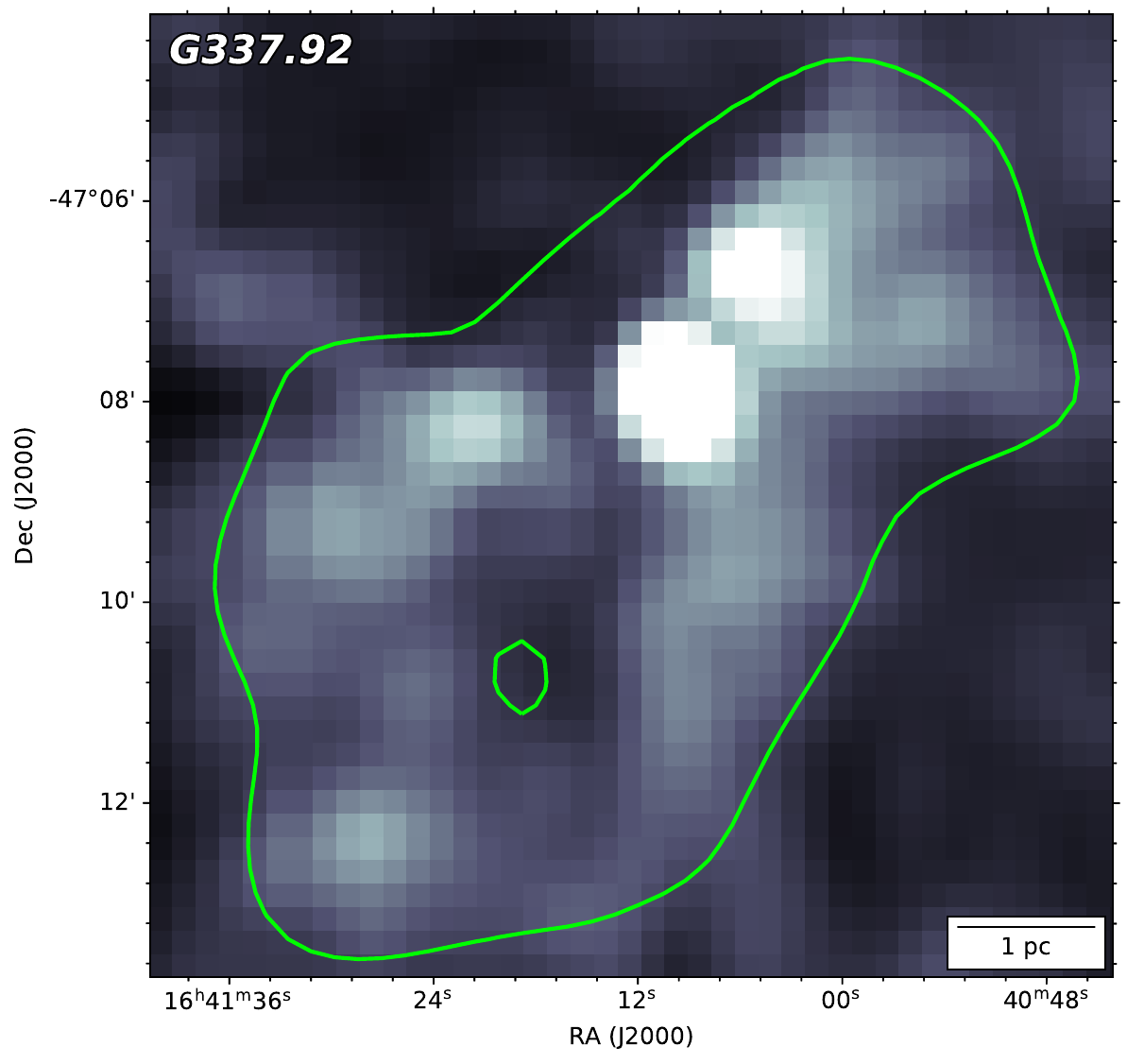} 
\hspace{-0.1cm}\includegraphics[scale=0.3]{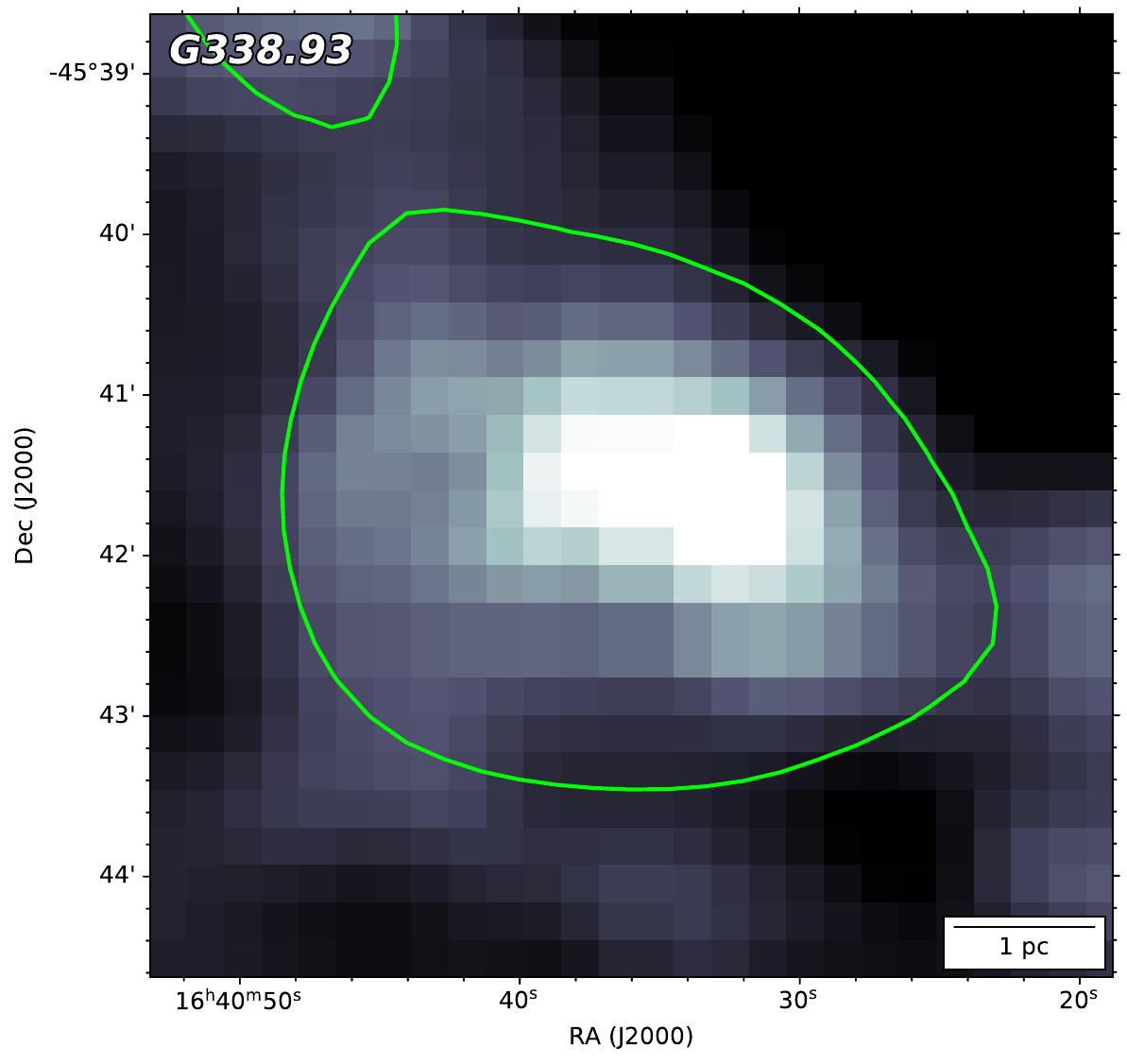} 
\hspace{-0.1cm}\includegraphics[scale=0.3]{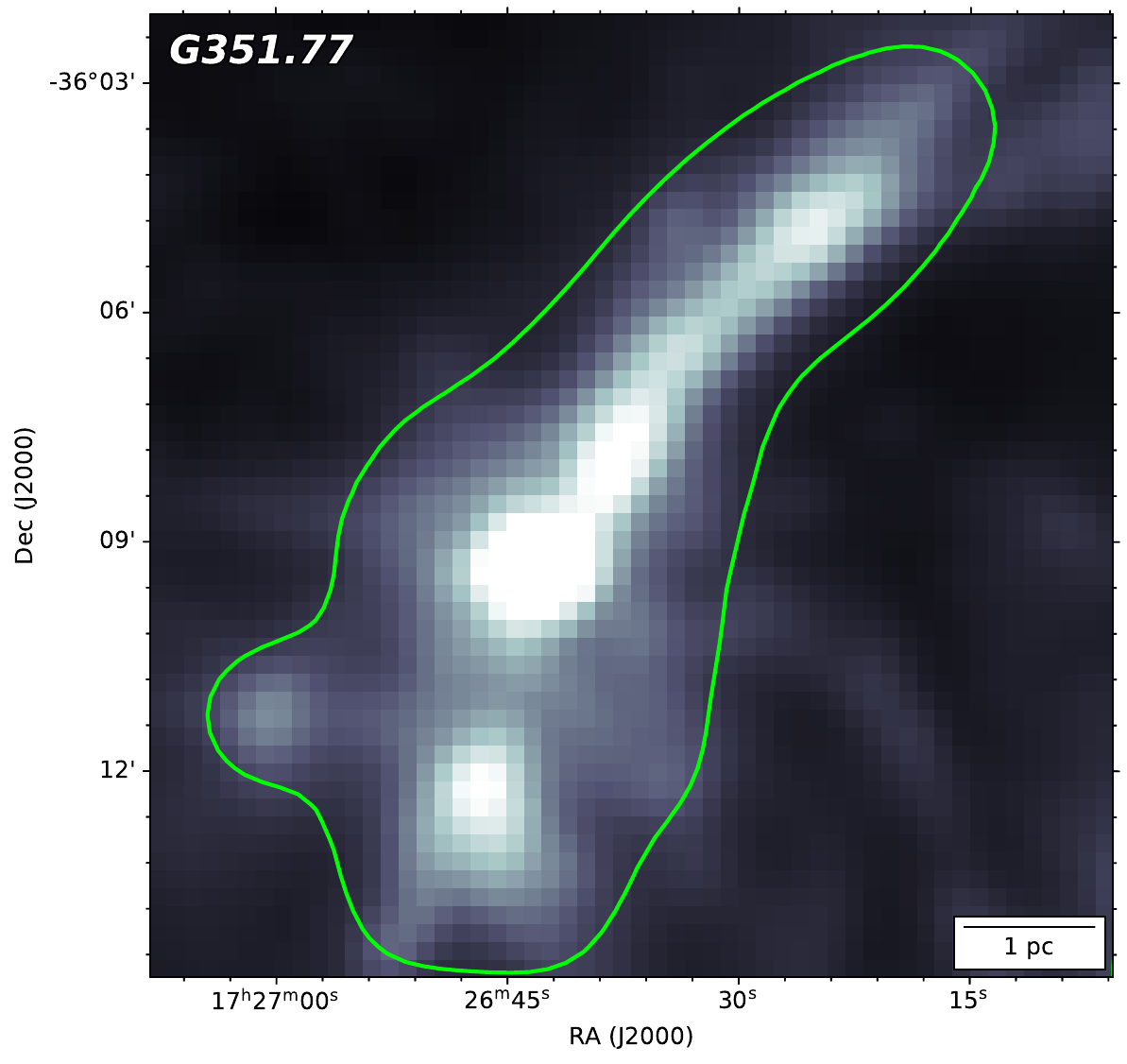}\\
\caption{
(Continued.)
}
\label{fig:nmaps_2}
\end{figure*}

\begin{figure*}[!ht]
\centering
\hspace{-0.3cm}\includegraphics[scale=0.3]{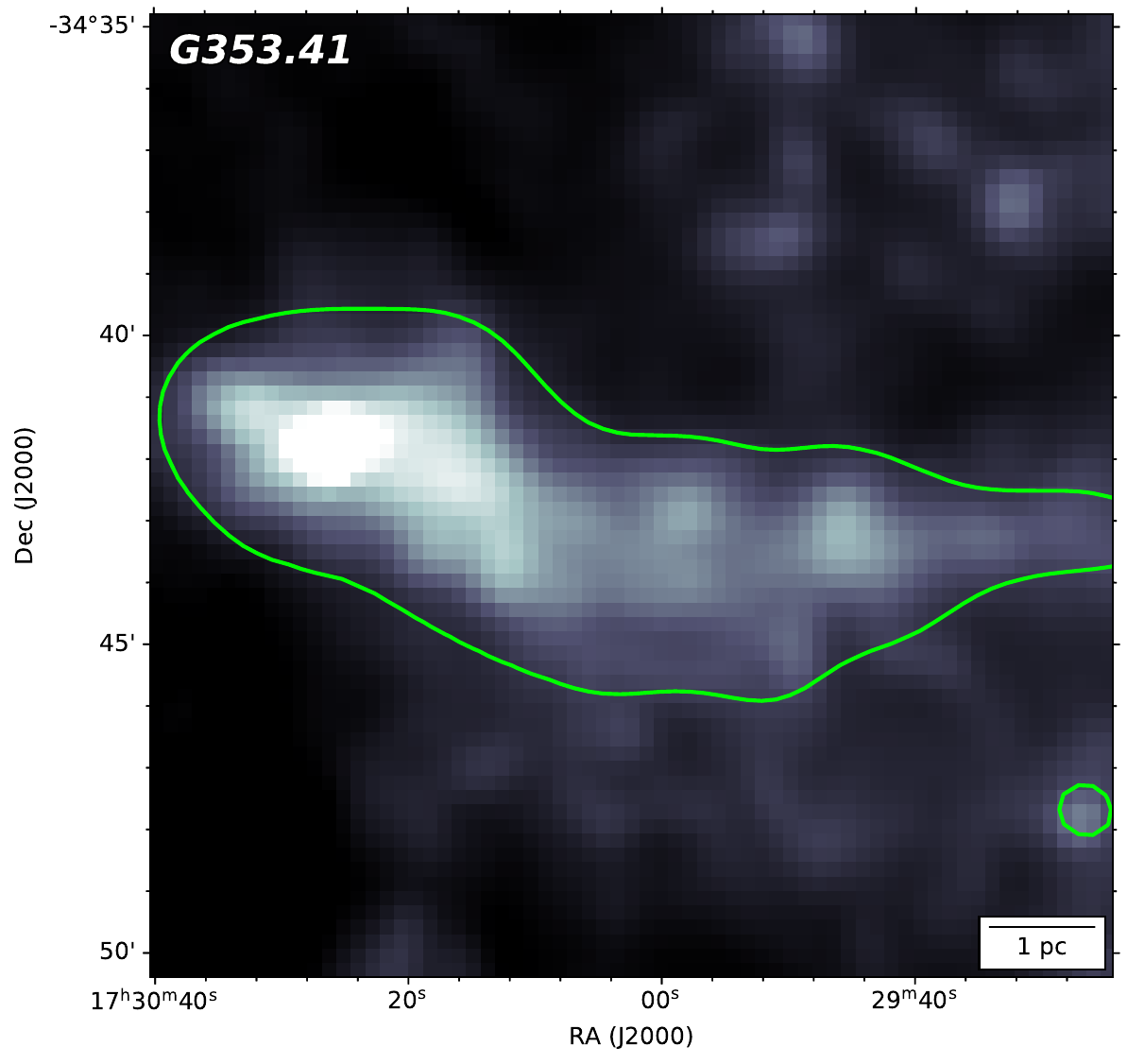} 
\hspace{-0.1cm}\includegraphics[scale=0.3]{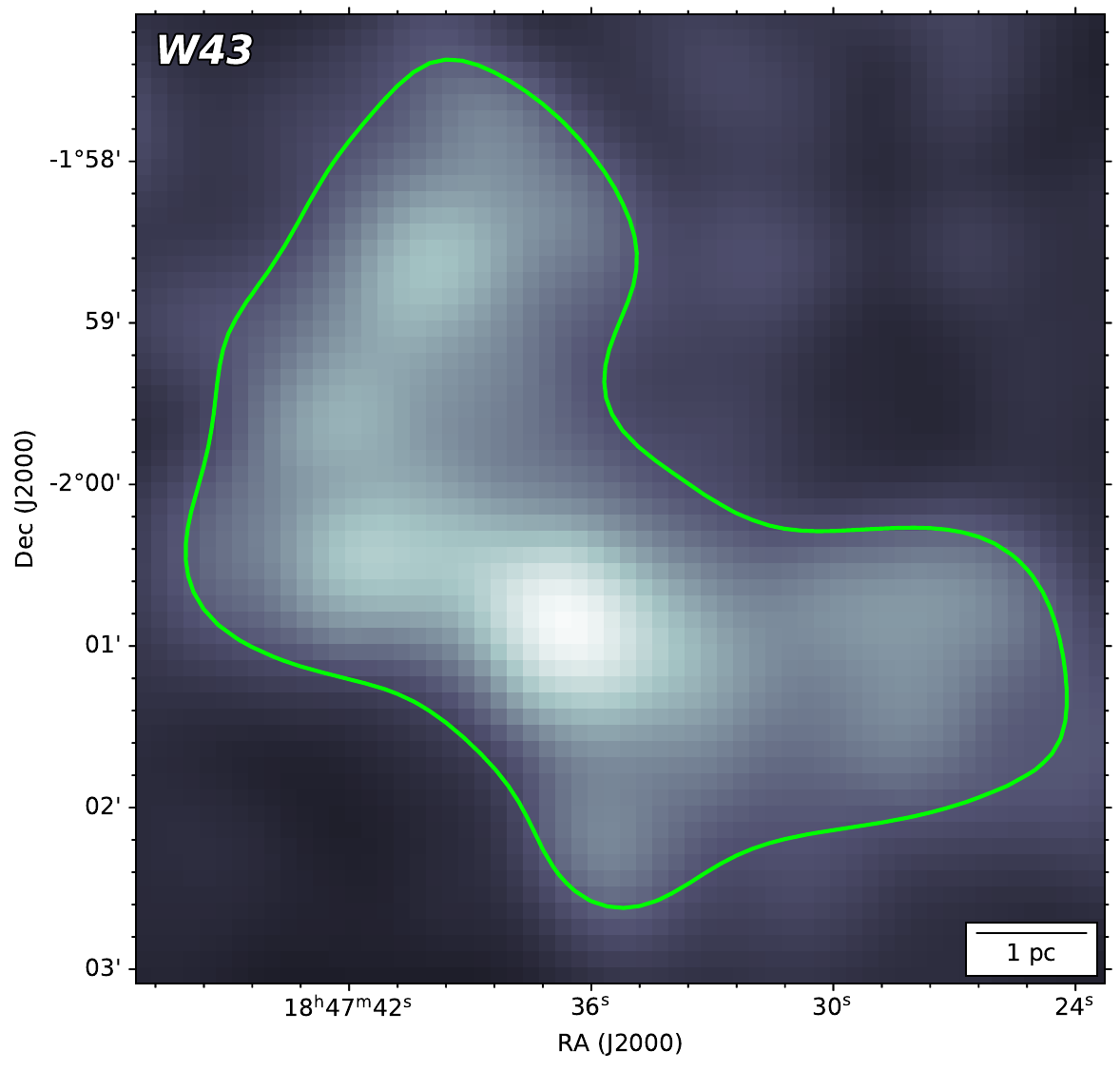} 
\hspace{-0.1cm}\includegraphics[scale=0.3]{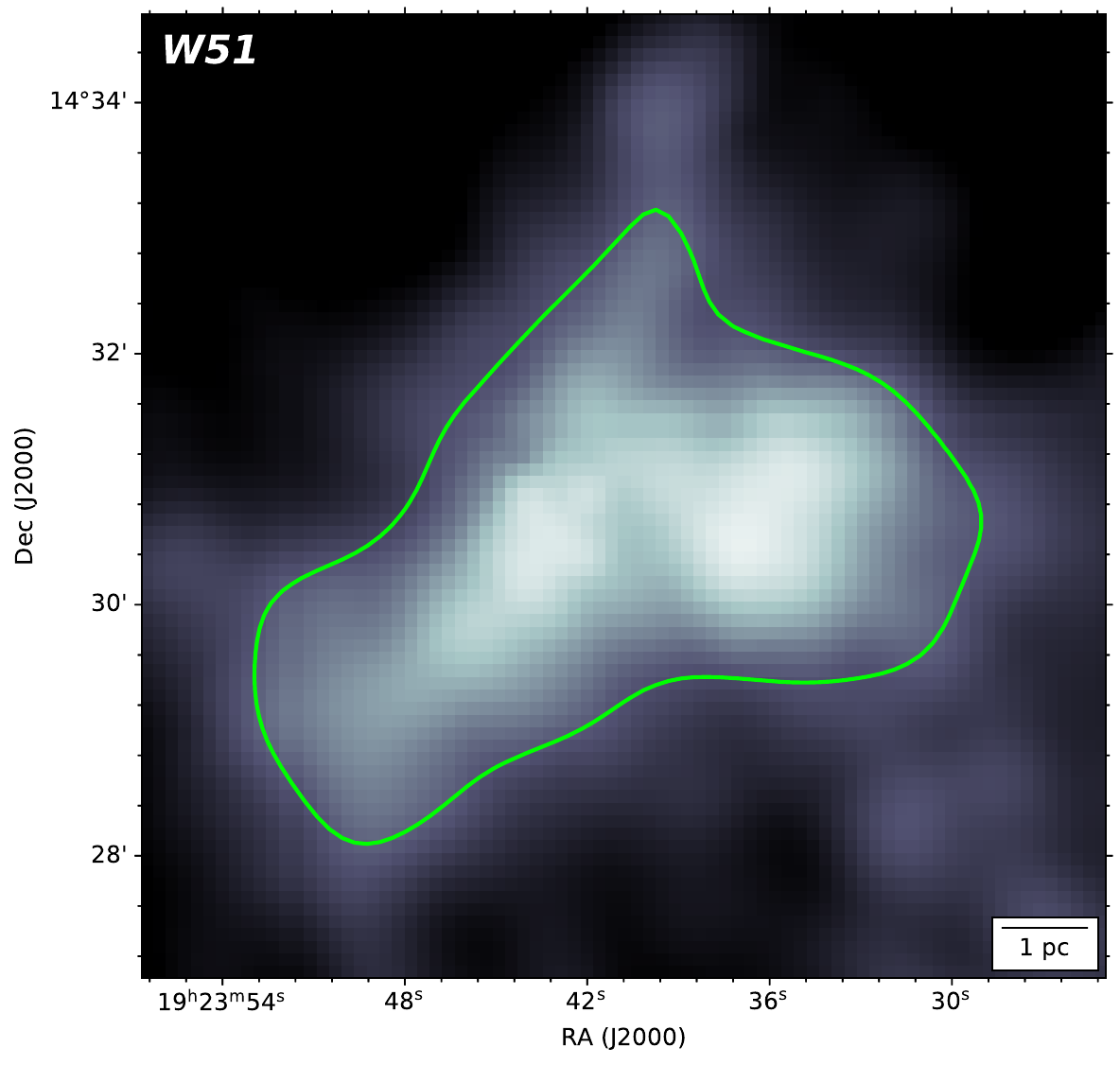}\\
\caption{
(Continued.)
}
\label{fig:nmaps_3}
\end{figure*}

During the least-square fitting procedure, the five data points are fitted with a preset weight, which is determined by the measured noise level throughout the image data. The pixelwise SED fitting results are column density (i.e. $N_{\rm\ssstyle H_2}$) maps for the nearby and distant clouds.
Figure \ref{fig:cloudcore}, Figure \ref{fig:nmaps_1}, Figure \ref{fig:nmaps_2}, and Figure \ref{fig:nmaps_3} show the threshold column density (lime contours) and the area associated with $M_{\rm gas}^{\rm bound}$ for each cloud.

\section{Core extraction} \label{app:sep}

In this section, we introduce the core extraction algorithm \texttt{SExtractor}\footnote{\href{https://sextractor.readthedocs.io}{https://sextractor.readthedocs.io/}} \citep{Bertin1996SExtractor}, as well as the input arguments in the procedure for core extraction in Section \ref{result:M_max}.

For each image, we first generate the background image as well as the root mean square (rms) noise, by iteratively clipping the local background histogram until convergence at $\pm3\sigma$ around its median \citep[see more details in][]{Bertin1996SExtractor}. The box is 256-pixel size which corresponds to 12'8, which means background emission is subtracted with more extended than 12'8. The background-subtracted images and noise rms maps serve as the input for extraction algorithm.

When running the algorithm, $\texttt{nthresh}=3$ is set to cut out the locally low SNR (\texttt{nthresh}$\times$rms) pixels. The deblending parameters (number of thresholds for deblending $\texttt{deblend\_nthresh}=512$, deblending contrast $\texttt{deblend\_cont}=10^{-5}$), and the parameter to control the minimum pixels in a core ($\texttt{min\_npix}$, that is, the number of pixels in a \textit{Herschel} or ALMA beam) are set for the source extraction procedure. 

As an example shown in Figure \ref{fig:cloudcore}, the orange ellipses indicate the extracted dense cores. The basic parameters include core location, angular size, and pixelwise integration. In the case of nearby clouds, since the column density ($N_{\rm \ssstyle H_2}$) map is given, the integration within the core mask is equivalent to the core mass as indicated by Equation \ref{eq:inte}. 
In the case of the ALMA-IMF intensity map, the integration gives the flux density emitted by the dust, which serves as the input in Equation \ref{eq:flux2mass}.

To obtain the physical size (deconvolved from the beam), we follow the method in \citet{Rosolowsky2010BGPSv1,Contreras2013CSC}, the deconvolved angular radius is written as,
\begin{equation}\label{eq:theta}
    \theta_{\rm deconv} = \eta\left[\left(\sigma^2_{\rm maj}-\sigma^2_{\rm bm}\right)\left(\sigma^2_{\rm min}-\sigma^2_{\rm bm}\right)\right]^{1/4},
\end{equation}
where $\sigma_{\rm maj}$ and $\sigma_{\rm min}$ are the dispersion of major and minor axes. The $\sigma_{\rm bm}$ is the averaged dispersion size of the beam (i.e., $\theta_{\rm bmaj}/\sqrt{8\ln2}$, where $\theta_{\rm bmaj}$ is the FWHM beam size). $\eta$ is a factor that relates the dispersion size of the emission distribution to the determined angular radius of the object. We have elected to use a value of $\eta=2.4$, which is the median value derived for a range of models consisting of a spherical, emissivity distribution \citep{Rosolowsky2010BGPSv1}. Therefore, the physical size of the core is derived from $R_{\rm deconv} = \theta_{\rm core} \times d$ ($d$ is the distance), which is listed in Table \ref{tab:mass}.

Orion A is located at a distance of 414 pc \citep{Menten2007A&A...474..515M}, which is 2--3 times farther than other nearby clouds. To identify cores with a consistent spatial resolution of $\sim$ 0.03 pc, we utilized the 10$''$ dust column density map of Orion A from \cite{Jiao2022SCPMA..6599511J} to perform the core extraction.

The final core catalog of nearby clouds is listed in Table \ref{tab:cat}.

We remark that the term, `core', has ambiguous meanings. 
It, to some extent, depends on the behavior of the core-identification algorithm(s) given specific image qualities. 
In the literature, the definition of `core' has systematically evolved with the angular resolutions of the major observing facilities.
Prior to the ALMA era, it was common to regard $\sim$0.1 pc scale gas over-densities as `core' \citep[e.g.,][]{Motte1998A&A...336..150M,Williams2000,Bergin2007ARA&A,Andre2010}. 
The core sizes we measured are consistent with those earlier studies.
In the more recent studies that were based on high-resolution ALMA data \citep[e.g., with $\lesssim$0.01 pc resolution; ][]{Cheng2024,Louvet2024,Molinari2025arXiv250305555M,Coletta2025arXiv250305663C}, `core' referred to gas over-densities with smaller spatial scales.
Nevertheless, these observations, with different resolutions (and the resulting core sizes), mostly report the CMF shapes which resembled the shape of the IMF. 
In addition, they also reported comparable power-law indices at the high-mass ends of their derived CMFs.

In our present work, based on the images with a 0.03~pc spatial resolution and using a specific core-identification algorithm, we found practical relations between $M_{\rm core}^{\rm max}$, $M_{\rm gas}^{\rm bound}$, and SFR.
Our results have no dependence on the core-identification in other studies.
Since the sizes of the compact structures we identified are systematically larger than the sizes of the `cores' report in the recent ALMA studies (e.g., \citealt{Cheng2024,Louvet2024,Molinari2025arXiv250305555M,Coletta2025arXiv250305663C}), one may wonder whether or not it is appropriate to refer to the compact structures we identified as `cores'.
We regard this as a terminology issue. 
For instance, we could have rename the compact structures we identified as `gas halos' just to distinguish them from the `cores' reported in the previous papers, and refer to the relation we identified as the $M_{\rm halo}^{\rm max}$-$M_{\rm gas}^{\rm bound}$ relation.
We do not find it helpful to create a new terminology when the definition of core remains ambiguous.
As shown in Figure \ref{fig:cmf}, the mass function of the structures we identified in nearby clouds closely resembles the canonical IMF/CMF. 
Therefore, we consider it appropriate to discuss these structures within the same conceptual framework as the 'cores' in previous studies.

\begin{deluxetable*}{ccccccccccc}[!ht]
\tablecaption{{Core catalog of nearby clouds}
\label{tab:cat}}
\centering
\tablehead{
\colhead{Cloud} & \colhead{ID} & \colhead{R.A.} & \colhead{Dec.} & \colhead{$\theta_{\rm maj}$} & \colhead{$\theta_{\rm min}$} & \colhead{PA} & \colhead{FWHM} & \colhead{$M_{\rm core}$} & \colhead{$N^{\rm peak} (\mathrm{H}_2$)} & \colhead{SNR}
}
\startdata
Aquila & 1 & 18:31:48.70 & -04:56:49.86 & 70.5 & 48.4 & -29.0 & 0.061 & 0.432 & 2.13e+21 & 6.7 \\
Aquila & 2 & 18:31:27.51 & -04:56:03.94 & 98.2 & 40.8 & -13.7 & 0.069 & 0.287 & 1.58e+21 & 4.6 \\
Aquila & 3 & 18:31:20.77 & -04:55:16.29 & 129.5 & 44.4 & -75.9 & 0.089 & 0.238 & 1.22e+21 & 4.0 \\
Aquila & 4 & 18:31:51.47 & -04:54:10.59 & 102.9 & 71.5 & -23.3 & 0.104 & 1.661 & 6.08e+21 & 20.6 \\
Aquila & 5 & 18:31:41.07 & -04:52:39.43 & 91.9 & 40.8 & 76.0 & 0.066 & 0.620 & 3.65e+21 & 10.4 \\
Aquila & 6 & 18:31:39.63 & -04:49:39.78 & 76.6 & 29.5 & 82.0 & 0.040 & 0.302 & 2.14e+21 & 6.5 \\
Aquila & 7 & 18:30:22.16 & -04:48:49.92 & 65.8 & 34.2 & -88.8 & 0.040 & 0.165 & 1.06e+21 & 4.2 \\
Aquila & 8 & 18:30:29.77 & -04:44:04.07 & 34.1 & 29.0 & -60.6 & \nodata & 0.075 & 9.14e+20 & 2.9 \\
Aquila & 9 & 18:32:01.66 & -04:43:19.26 & 76.6 & 37.3 & 2.1 & 0.052 & 0.340 & 1.87e+21 & 6.4 \\
Aquila & 10 & 18:32:23.12 & -04:42:15.14 & 37.9 & 35.9 & 43.6 & \nodata & 0.273 & 3.22e+21 & 12.5 \\
\enddata
\tablecomments{The properties of cores identified in nearby clouds are listed.
Here, only the top ten rows of Aquila are shown.
The complete table will be available online. 
}
\end{deluxetable*}

\section{Temperature estimation of cores within massive clouds} \label{app:temp}

\begin{figure*}
\centering
\includegraphics[width=0.49\linewidth]{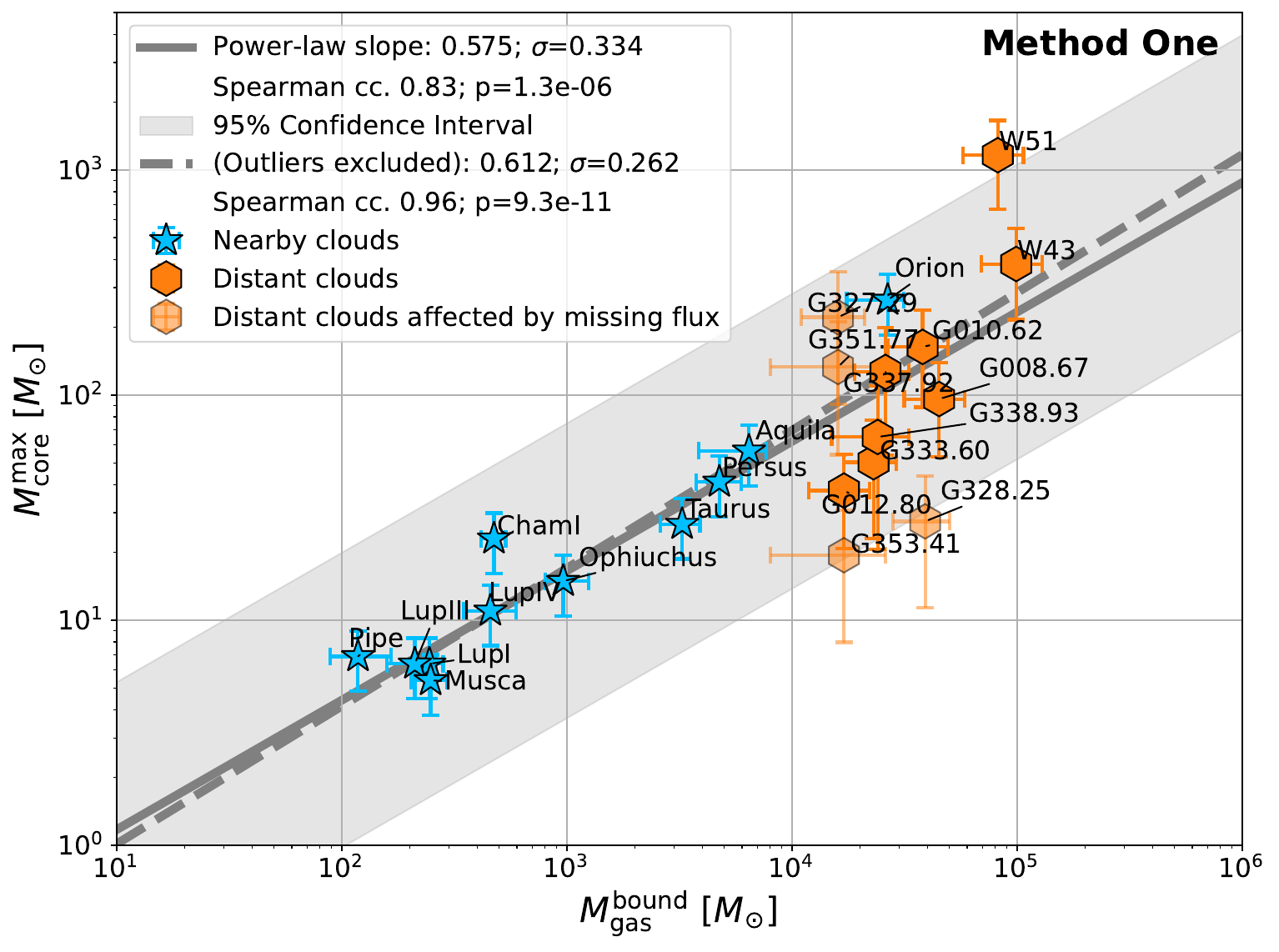}
\includegraphics[width=0.49\linewidth]{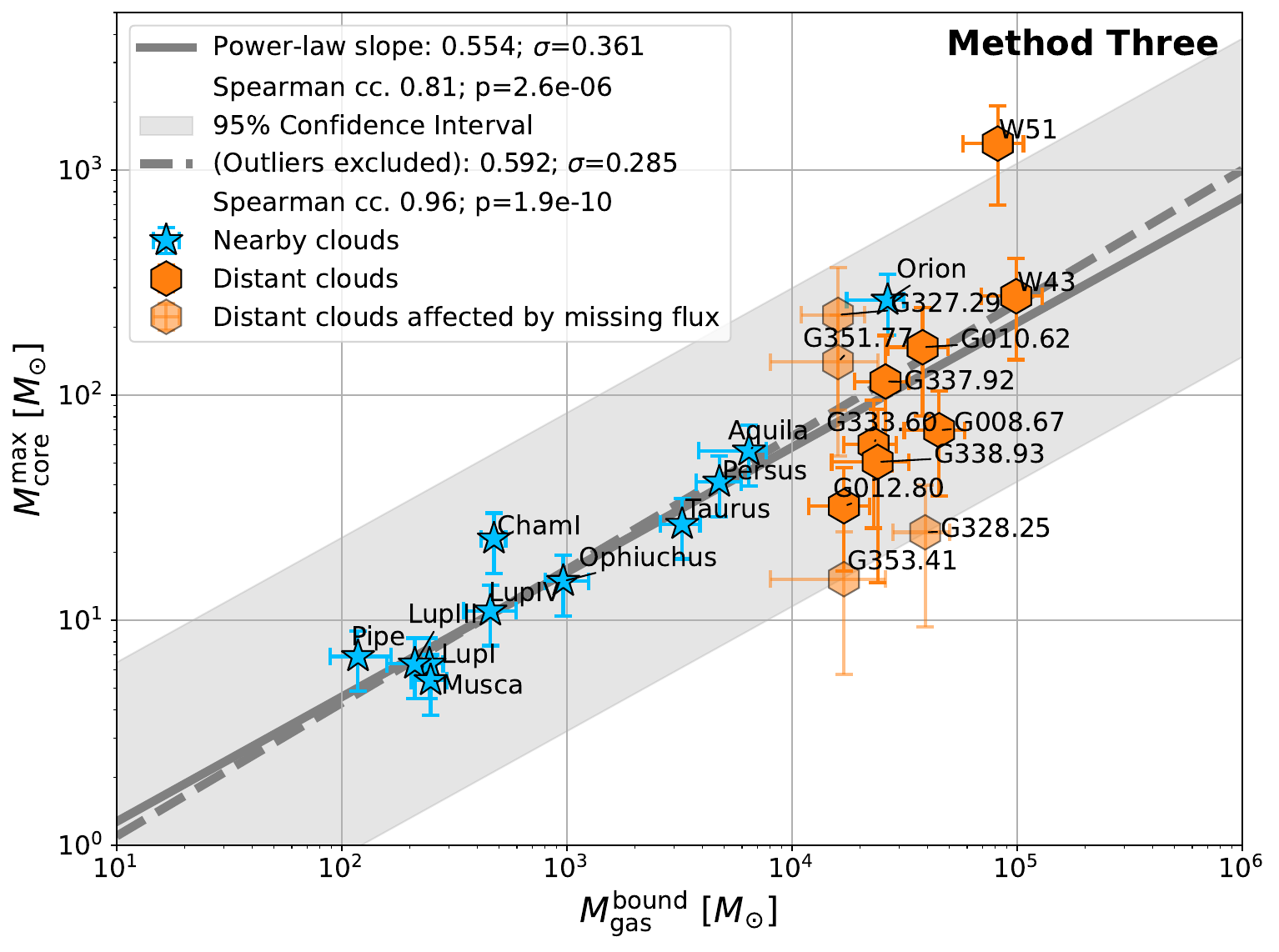}
\caption{The most massive core mass $M^{\rm max}_{\rm core}$ versus its parental gravitational bound gas $M_{\rm gas}^{\rm bound}$, using the methods one and three. 
The method two is adopted for Figure \ref{fig:M-M}, so it is not shown in this figure. The blue stars indicate the nearby clouds and the orange hexagons indicate the distant massive clouds. Using the other two different temperature estimation methods, $M^{\rm max}_{\rm core}$ have two versions for the power-law fitting. 
The power-law index is listed with a $1\sigma$ to indicate how the data scatters. The Spearman correlation coefficient (cc.) and its p-value are to show the significance of the fitting. The $1\sigma$ data scatter and the gray-shaded region correspond to the regimes with 95\% confidence interval. \label{fig:test}}
\end{figure*}

We tried three methods of temperature estimation for these massive dense cores, which are listed below.

$\bullet$ Method One. We cross-match the sources with the highest flux density (potentially the most massive one) with the \textit{Herschel} PACS 70 $\mu$m source catalog \citep{Molinari2016HiGAL}. Considering the effects of optical depth, the internal bolometric luminosity can be obtained from the 70~$\mu$m flux density \citep{Elia2017HiGAL},
\begin{equation}
L_{\mathrm{int}} = 25.6\left(\frac{S_{70\mu\mathrm{m}}}{10\, \mathrm{Jy}}\right) \left(\frac{d}{1\, \mathrm{kpc}}\right)^2 \left(\frac{\tau_\nu}{1-e^{-\tau_\nu}}\right) \, L_{\odot}
\end{equation}
where $S_{70\mu\mathrm{m}}$ is the integrated 70 $\mu$m flux density of the source, $d$ is the distance to the clump, and $\tau$ is the dust opacity at 70~$\mu$m. $\tau_{\rm dust} = \kappa \Sigma_{\rm dust}$ where $\kappa\simeq100$~cm$^2$/g at 70~$\mu$m \citep{Ossenkopf1994} and $\Sigma_{\rm dust}$ is typically order of 0.01~g cm$^{-2}$ for massive cores at the scale of 2000-6000~AU \citep{Motte2018NatAs...2..478M,Xu2024ASSEMBLE}.

Now we calculate the mean mass-weighted temperature $T^{\rm mean}_{\rm dust}$ within core radius $r$ \citep{Terebey1993},
\begin{equation}
	T^{\rm mean}_{\rm dust} = \frac{3}{2} T_0 \left(\frac{L_\mathrm{int}}{L_0}\right)^{1/6} \left(\frac{r}{r_0}\right)^{-1/3}
\end{equation}
where $L_\mathrm{int}$ is the internal luminosity and $r$ is the core's radius which can be obtained from the core extraction algorithm (see Equation \ref{eq:theta} in Appendix \ref{app:sep}). The reference values are $T_0 = 25$\,K, $L_0 = 520$~$L_{\odot}$, and $r_0=0.032$\,pc. The scaling relation above assumes dust opacity index $\beta=2$ and density profile follows $\rho(r)\propto r^{-2}$. 

$\bullet$ Method Two. All of these massive dense cores are showing hot molecular species CH$_3$OCHO line emission \citep{Bonfand2024}, which means there should be internal heating sources like OB stars. So in the method two, we simply assume CH$_3$OCHO-bearing cores have a uniform temperature of 100 K. \citet[see their Section 5.2]{Bonfand2024} also argue that some sources like W51-E, W51-IRS2, and G327.29 should contain temperature as high as 300 K. For example, \citet{Ginsburg2017} report a peak excitation temperature $>350$ K based on the analysis of CH$_3$OH emission lines at 1800-au resolution in W51-E. So we use the same convention of 300 K for the case of W51-E, W51-IRS2, and G327.29, as what \citet{Bonfand2024} did. 

$\bullet$ Method Three. Based on the reason discussed in Method Two, we also test the case where the temperature is also set to 100 K without any exception. 

In Figure \ref{fig:test}, we present the data points and fitting results between the most massive core mass $M^{\rm max}_{\rm core}$ versus its parental gravitational bound gas $M_{\rm gas}^{\rm bound}$ for the methods one and three. 
Method two is adopted in Figure \ref{fig:M-M} for its least data scatter. 
So it is not shown here to avoid duplication. Generally speaking, the slope is around 0.5--0.6 with a systematic uncertainty of 0.07 from method choice. 
We also list the Spearman correlation coefficient and its corresponding p-value in each panel. 
In all of the three cases, the correlation is significant.

\begin{deluxetable*}{c D@{+/-}D D@{+/-}D D@{+/-}D D@{+/-}D D@{+/-}D D@{+/-}D}
\tablecaption{Parameters the most massive cores from three methods}
\tablehead{
\colhead{ALMA-IMF Field} & \multicolumn{8}{c}{Method One} & \multicolumn{8}{c}{Method Two} & \multicolumn{8}{c}{Method Three}\\
\cmidrule(r){2-9} \cmidrule(r){10-17} \cmidrule(r){18-25} 
 & \multicolumn{4}{c}{T$_{\rm dust}$ (K)} & \multicolumn{4}{c}{Mass ($M_{\odot}$)} & \multicolumn{4}{c}{T$_{\rm dust}$ (K)} & \multicolumn{4}{c}{Mass ($M_{\odot}$)} & \multicolumn{4}{c}{T$_{\rm dust}$ (K)} & \multicolumn{4}{c}{Mass ($M_{\odot}$)}
}
\decimals
\startdata
G008.67 & 69.7 & 2.1 & 101.8 & 30.7 & 100.0 & 16.0 & 70.1 & 23.2 & 100.0 & 16.0 & 69.9 & 23.2 \\
G010.62 & 91.4 & 3.1 & 177.3 & 56.4 & 100.0 & 16.0 & 163.8 & 55.9 & 100.0 & 16.0 & 163.4 & 55.8 \\
G012.80 & 80.0 & 2.4 & 39.9 & 11.9 & 100.0 & 16.0 & 32.1 & 10.6 & 100.0 & 16.0 & 32.0 & 10.5 \\
G327.29 & 86.9 & 5.9 & 264.9 & 111.7 & 300.0 & 32.0 & 67.1 & 28.5 & 100.0 & 16.0 & 226.2 & 98.1 \\
G328.25 & 86.0 & 5.8 & 28.5 & 12.0 & 100.0 & 16.0 & 24.6 & 10.7 & 100.0 & 16.0 & 24.5 & 10.7 \\
G333.60 & 111.2 & 6.3 & 53.0 & 20.3 & 100.0 & 16.0 & 60.2 & 24.0 & 100.0 & 16.0 & 60.2 & 23.9 \\
G337.92 & 84.8 & 5.3 & 136.0 & 54.9 & 100.0 & 16.0 & 115.1 & 48.4 & 100.0 & 16.0 & 115.3 & 48.4 \\
G338.93 & 75.4 & 6.5 & 68.1 & 33.1 & 100.0 & 16.0 & 50.6 & 25.1 & 100.0 & 16.0 & 50.4 & 25.1 \\
G351.77 & 92.5 & 6.2 & 151.5 & 64.0 & 100.0 & 16.0 & 140.4 & 60.9 & 100.0 & 16.0 & 141.1 & 61.6 \\
G353.41 & 76.6 & 5.2 & 20.1 & 8.5 & 100.0 & 16.0 & 15.3 & 6.6 & 100.0 & 16.0 & 15.3 & 6.7 \\
W43-MM1 & 72.6 & 1.9 & 394.7 & 115.3 & 100.0 & 16.0 & 275.0 & 88.3 & 100.0 & 16.0 & 274.0 & 88.1 \\
W43-MM2 & 60.4 & 1.5 & 399.8 & 116.1 & 100.0 & 16.0 & 228.5 & 72.9 & 100.0 & 16.0 & 228.2 & 73.3 \\
W43-MM3 & 58.7 & 1.5 & 235.6 & 68.5 & 100.0 & 16.0 & 134.5 & 43.6 & 100.0 & 16.0 & 134.3 & 43.2 \\
W51-E & 98.9 & 1.9 & 1222.7 & 342.8 & 300.0 & 32.0 & 323.8 & 94.9 & 100.0 & 16.0 & 1223.7 & 381.5 \\
W51-IRS2 & 97.0 & 2.0 & 1344.5 & 376.1 & 300.0 & 32.0 & 321.8 & 94.0 & 100.0 & 16.0 & 1308.8 & 408.1
\enddata
\tablecomments{The dust temperature estimation and the most massive core mass in three methods. $1\sigma$ uncertainty is listed for uncertainty calculation. \label{tab:method}}
\end{deluxetable*}


\section{Comparison between SFR$^{\rm theory}$ and SFR$^{\rm observation}$} \label{app:sfr}

\begin{figure}
\includegraphics[height=6.5cm]{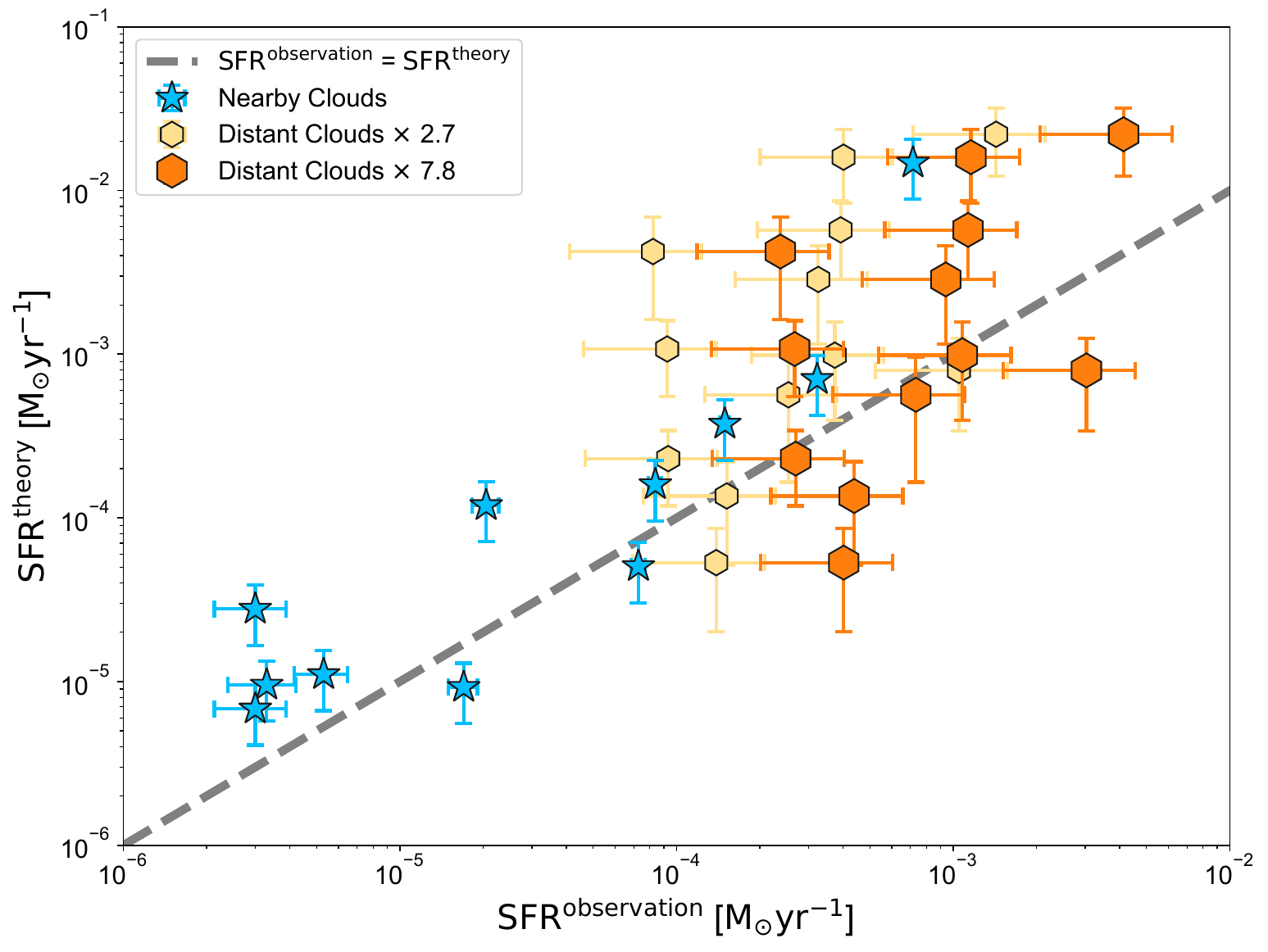}
\caption{
Comparison between SFR$^{\rm observation}$ and SFR$^{\rm theory}$.
The black dashed line indicates SFR$^{\rm observation}=$SFR$^{\rm theory}$.
}\label{fig:sfr}
\end{figure}

In Section \ref{discuss:sf_law}, we derived the SFRs$^{\rm theory}$ for the target clouds and compared them with the Gao-Solomon relation.
This section extends the comparison to include the observationally constrained SFR$^{\rm observation}$.

For nearby clouds, the SFR$^{\rm observation}$ is constrained by counting young stellar objects (YSOs; \citealt{Evans2009ApJS..181..321E,Heiderman2010}).
For distant clouds, the SFR is constrained by the total infrared luminosities.
The total infrared luminosity is calculated according to the prescription in \citet{Sanders1996ARA&A..34..749S} based on the four {\it IRAS} flux bands observations.
SFR$^{\rm observation}$ of distant clouds are calculated by the relation, SFR $\approx 2 \times 10^{-10}$ ($L_{\rm IR}$/$L_{\odot}$) $M_{\odot}$ yr$^{-1}$, following \citet{Kennicutt1998,Gao2004,Lada2012}.
However, as pointed out by \cite{Chomiuk2011AJ....142..197C} and \cite{Lada2012}, the SFRs from far-infrared observations require upward adjustments to match those from the YSO counting method.
We tried adjusting the SFR$^{\rm observation}$ upward by a factor of 2.7 as suggested by \cite{Lada2012}, and a factor of 7.8 based on the measurements towards Orion A and B clouds \citep[][and Jiao et al. submitted]{Lada2012}.
Figure \ref{fig:sfr} shows the comparison between SFR$^{\rm theory}$ and SFR$^{\rm observation}$.
We found that, while there is broad agreement between SFR$^{\rm theory}$ and SFR$^{\rm observation}$, nearby clouds closely follow the line of equality, whereas distant clouds deviate significantly.
Since the SFR$^{\rm observation}$ of the distant clouds sample have large intrinsic scattering, the upward adjustments with the two different factors did not change this conclusion.

\bibliographystyle{yahapj}
\bibliography{references}
\end{document}